\documentclass[aps,pre,twocolumn,superscriptaddress]{revtex4}

\usepackage{graphicx}
\usepackage{amsmath}
\usepackage{amssymb}
\usepackage{color}
\usepackage{amscd}
\usepackage{bm}
\usepackage{subfigure}
\usepackage{psfrag}

\begin{document}
\title{Motor-driven Dynamics of Cytoskeletal Filaments in Motility Assays}

\author{Shiladitya Banerjee}
\affiliation{Physics Department, Syracuse University, Syracuse, NY 13244, USA}

\author{M. Cristina Marchetti}
\affiliation{Physics Department \& Syracuse Biomaterials Institute, 
Syracuse University, Syracuse, NY 13244, USA}

\author{Kristian M\"uller-Nedebock}
\affiliation{Institute of Theoretical Physics/Department of Physics, 
Stellenbosch University, Matieland 7602, South Africa}

\date{\today}

\begin{abstract}
We model analytically the dynamics of a cytoskeletal filament in a motility assay.   The filament is described as  rigid rod free to slide in two dimensions. The motor proteins consist of   polymeric tails  tethered to the plane and modeled as linear springs and motor heads that  bind to the filament.   As in related models of rigid and soft two-state motors, the  binding/unbinding dynamics of the motor heads and the dependence of the transition rates on the load exerted by the motor tails play a crucial role in controlling the filament's dynamics. Our work shows that the filament effectively behaves as a self-propelled rod at long times, but with non-Markovian noise sources arising from the coupling to the motor binding/unbinding dynamics. The effective propulsion force of the filament and the active renormalization of the various friction and diffusion constants are calculated in terms of microscopic motor and filament parameters.  These quantities could be probed by optical force microscopy.

\end{abstract}

\maketitle

There has recently been renewed interest in motility assays
where semiflexible actin filaments are driven to slide over a ``bed" of
myosin molecular motors. Recent experiments at high actin density have
revealed that the collective behavior of this simple active system is
very rich, with propagating density waves and large scale-swirling
motion~\cite{Schaller2010,Butt2010}, not unlike those observed in dense bacterial suspensions~\cite{Weibel2009}. In
an actin  motility assay the polymeric tails of myosin motor proteins
are anchored to a surface, while their heads can bind to actin
filaments~\cite{Riveline1998}.
Once bound, the motor head  exerts forces and drives the filament's
motion. 
This system provides possibly the simplest realization of an active system that allows detailed semi-microscopic modeling.

Stochastic models of the collective action of motor proteins on cytoskeletal filaments in \emph{one dimension} have been considered before by several authors, with emphasis on the acto-myosin system in muscles and on the mitotic spindle~\cite{GuerinReview2010}.  When working against an elastic load,  the motor assemblies have been shown to drive periodic spontaneous activity in the form of oscillatory instabilities, which in turn have been observed ubiquitously in a variety of biological systems~\cite{JulicherProst1997,Grill2005,GuntherKruse2007,VilfanFrey2005,CamaletJulicher2000}. These instabilities arise in the model from the collective action of the motors and the breaking of detailed balance in their dynamics and manifest themselves as a negative effective friction of the filament.  When free to slide under the action of an external force, the filament can exhibit bistability that manifests itself as hysteresis in the force velocity-curve~\cite{JulicherProst1995,Badoual2002}. A large body of earlier work has modeled the motors as rigid two-state systems  attached to a backbone and bound by the periodic potential exerted by the filament on the motor head~\cite{JulicherProst1995,JulicherProst1997,Placais2009}. In a second class of models the motors have been modeled  as flexible springs~\cite{Gibbons2001,Kraikivski2006}. The motor heads
 bind to the filament and unbind at a load-dependent rate. In this case the dynamic instability arises from the dependence of the unbinding rate on the tension exerted by springs
~\cite{Brokaw1975,Vilfan1999,Kafri2009}. Recent work by Gu{\'e}rin et al.~\cite{Guerin2010} has generalized the two-state model by taking into account the flexibility of the motors, showing that both models can be obtained in a unified manner for different values of a parameter that compares
the stiffness of the motors to the stiffness of the periodic potential provided by the filament.

In this paper we consider a  model of a rigid filament free to slide in \emph{two dimensions} under the driving action of motor proteins uniformly tethered to a two-dimensional plane. The model considered is a modification of the ``crossbridge" model first introduced by Huxley in 1957 to describe motor-induced contractile behavior of muscle fibers~\cite{Huxley1957}.   The motor proteins'  polymeric tails  are modeled as linear springs that pull back on the bound motor heads. After attachment, the motor heads slide along the filament at a velocity that depends on the load exerted by the flexible motor tails. The sliding and subsequent detachment play the role of the motor's power stroke. The  binding/unbinding dynamics of the motor heads and the dependence of the transition rates on the load exerted by the motor tails play a crucial role in controlling the dynamics of the fie, effectively yielding  non-Markovian noise sources on the filament. 
Related models have been studied numerically~\cite{Gibbons2001,Kraikivski2006,Vilfan2009}.
The results presented here are obtained by generalizing  to two dimensions the mean field approximation for the motor dynamics described for instance in Ref.~\cite{Grill2005}. The mean-field theory neglects convective nonlinearities in the equation for the probability of bound motors and correlations in the motors on/off dynamics, but it is expected to be adequate on time scales large compared to that of the motor on/off dynamics  and for a large number of motors. This is supported by the results of ~\cite{JulicherProst1995}  for a model of rigid two-state motors.

We begin by revisiting the one-dimensional problem. We discuss the steady-state response of the filament to an external force and present new results on the dynamics of fluctuations about the sliding steady state. The force-velocity curve is  evaluated analytically and exhibits bistability and hysteresis, as obtained in Ref.~\cite{JulicherProst1995} for a rigid two-state motor model. A new result is an expression for the effective propulsion force on the filament due to the motors in terms of physical parameters characterizing the motor proteins. Next, we analyze the fluctuations about the steady state  by evaluating the mean-square displacement of the filament. We show that the coupling to the motor binding/unbinding dynamics yields non-Markovian noise sources with time correlations controlled by the duration of the motors' binding/unbindig cycle. Since the filament has a finite motor-induced velocity even in the absence of applied force, the mean-square displacement is ballistic at long time. The fluctuations of displacement about this sliding state are, however, diffusive at long times with an enhanced diffusion constant. This enhancement is controlled by the dependence of the motors' unbinding rate on the load exerted on the bound motors' heads by the tethered tails and vanishes for unloaded motors. 

We then consider the case of a filament in two dimensions, to analyze the effect of the coupling of translational and rotational degrees of freedom in controlling the dynamics. At steady state, motors yield an effective propulsion force along the long axis of the filament, as in one dimension, but no effective torque. This is in contrast to phenomenological models considered in the literature~\cite{Teeffelen2008} that have considered the dynamics of active rod-like particles in the presence of both effective internal forces and torques. As a result, in the steady-state  the filament slides along its long axis and the dynamics in this direction is essentially one dimensional, with a motor-induced negative friction instability and bistability and hysteresis in the response to an external force. Motors do enhance both the transverse and the rotational friction coefficients of the filament. The enhancement of rotational friction could be probed by measuring the response to an external torque. Since the finite motor-induced propulsion is along the filament axis, whose direction is in turn randomized by rotational diffusion, the mean velocity of the filament  is zero in the absence of external force, unlike in the one-dimensional case. The mean square displacement is therefore diffusive at long times, with behavior controlled by the interplay of non-Markovian effects due to the coupling to motor dynamics with coupled translational and rotational diffusions. The filament performs a persistent random walk that consists of ballistic excursions at the motor-induced propulsion speed, randomized by both rotational diffusion and the motor binding/undinding dynamics. The crossover to the long-time diffusive behavior is controlled by the interplay of motor-renormalized diffusion rate and duration of the motor binding/unbinding cycle.  The effective diffusion constant is calculated in terms of microscopic motor and filament parameters. Its dependence on activity, as characterized by the rate of ATP consumption, could be probed in actin assays.

Finally, our work provides a microscopic justification of a simple model used in the literature~\cite{BaskaranMarchetti2008} that describes a cytoskeletal filament interacting with motor proteins tethered to a plane as a ``self-propelled" rod, although it also shows that the effective noise is rendered non-Markovian by the coupling to the motors' binding/unbing dynamics.  It also provides microscopic expressions for the self-propulsion force and the various friction coefficients in terms of motor and filament parameters and shows that this effective model fails beyond a critical value of motor activity, where the effective friction changes sign and the filament exhibits bistability and hysteresis.

\section{The Model}
\label{Sec:Model}

In our model the motor proteins are described as composed of  polymeric tails
attached permanently to a two-dimensional fixed substrate  and motor heads that
can bind reversibly to the filament.  Once bound, a motor head moves along the
filament thereby stretching the tail.  This gives rise to a load force on the
motor head and on the filament.  Eventually excessive load leads to detachment
of the motor head.

\subsection{Filament dynamics} 
\label{Sec:FilamentDyn}

The actin filament is modeled as a rigid polar rod of  length $L$ that can
slide in two dimensions. It is described  by the position ${\bf r}$ of its
center of mass and a unit vector ${\bf
\hat{u}}=\left(\cos(\theta),\sin(\theta)\right)$ directed along the rod's
long axis \emph{away } from the polar direction of the rod, which is in turn
defined as the direction of motion of bound motors. In other words, bound motors
move along the rod in the direction $-{\bf \hat{u}}$.   In contrast to most
previous work~\cite{JulicherProst1997,Grill2005,Placais2009,Guerin2010}, 
and given our interest in modeling actin motility assays, we
assume the substrate is fixed and consider the dynamics of the filament.  Our
goal is to understand the role of the cooperative driving by motors in
controlling the coupled rotational and translational
dynamics of the rod. 

The dynamics of the filament is described by coupled equations
for the translational and orientational degrees of freedom, given by
\begin{subequations}
\begin{gather} \label{r-eq} {\bf
\underline{\underline{\bm\zeta}}}\cdot\partial_t{\bf r}= {\bf F}_\text{a} + 
{\bf F}_\text{ext}+\bm\eta(t)\;,\\
\label{theta-eq}
\zeta_{\theta} \partial_t \theta = T_\text{a} +T_\text{ext} + \eta_\theta(t)\;.
\end{gather}
\end{subequations}
Here we have grouped the forces and torques into the effects due to the motors,
\emph{i.e.} the activity, $\bf{F}_\text{a}$ and $T_\text{a}$, external
forces and torques $\bf{F}_\text{ext}$ an $T_\text{ext}$ and the stochastic
noise not due to motors.  The friction tensor is given by ${\bf
\underline{\underline{\bm\zeta}}}=\zeta_{\Vert} {\bf \hat{u}\hat{u}} +
\zeta_{\perp}\left({\bf
\underline{\underline{\bm\delta}}-\hat{u}\hat{u}}\right)$ with $\zeta_\Vert$
and $\zeta_\perp$ the friction coefficients for motion longitudinal and
transverse to the long direction of the rod, and $\zeta_\theta$ is the
rotational friction coefficient.  For the case of a long, thin rod of interest
here, $\zeta_\Vert=\zeta_\perp/2$. The random force $\bm\eta(t)$ and random
torque $\eta_\theta(t)$ describe noise in the system, including
nonthermal noise sources.  For simplicity we assume 
that both $\bm\eta(t)$ and
$\eta_\theta(t)$ describe Gaussian white noise, with zero mean and correlations
$\langle\eta_i(t) \eta_j(t') \rangle = 2 B_{ij}\delta(t-t')$ and
$\langle\eta_\theta(t) \eta_\theta(t') \rangle = 2 B_{\theta}\delta(t-t')$, where 
$B_{ij}=B_{\Vert}  \hat{u}_i\hat{u}_j +
B_{\perp}\left(\delta_{ij}-\hat{u}_i\hat{u}_i\right)$.

\subsection{Individual motor dynamics}
\label{Sec:MotorDyn}

\begin{figure}
\begin{center}
\includegraphics[width=7cm]{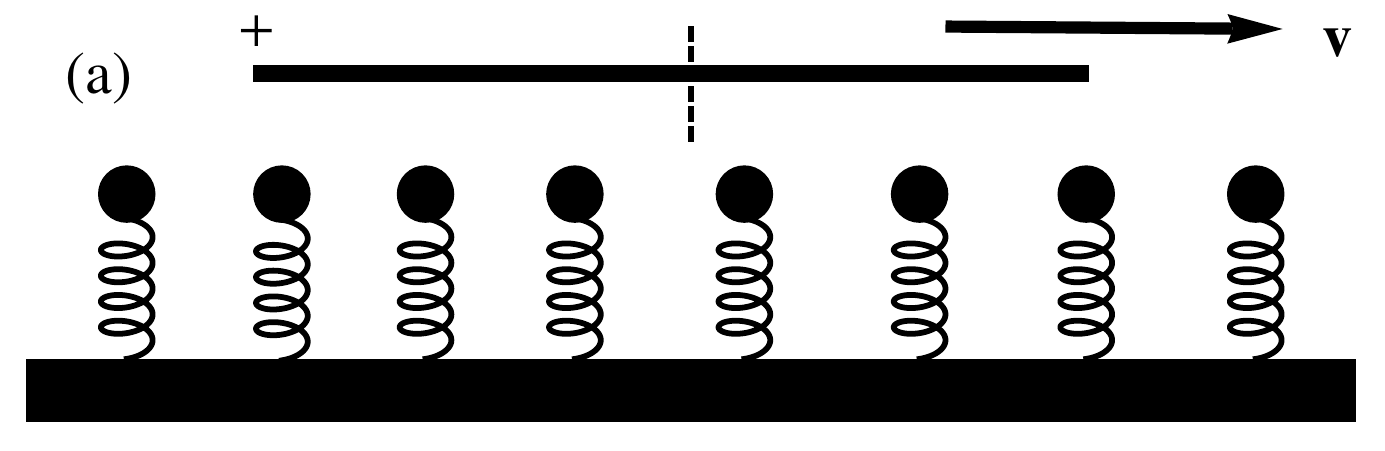}\vspace{0.2in}\\
\includegraphics[width=7cm]{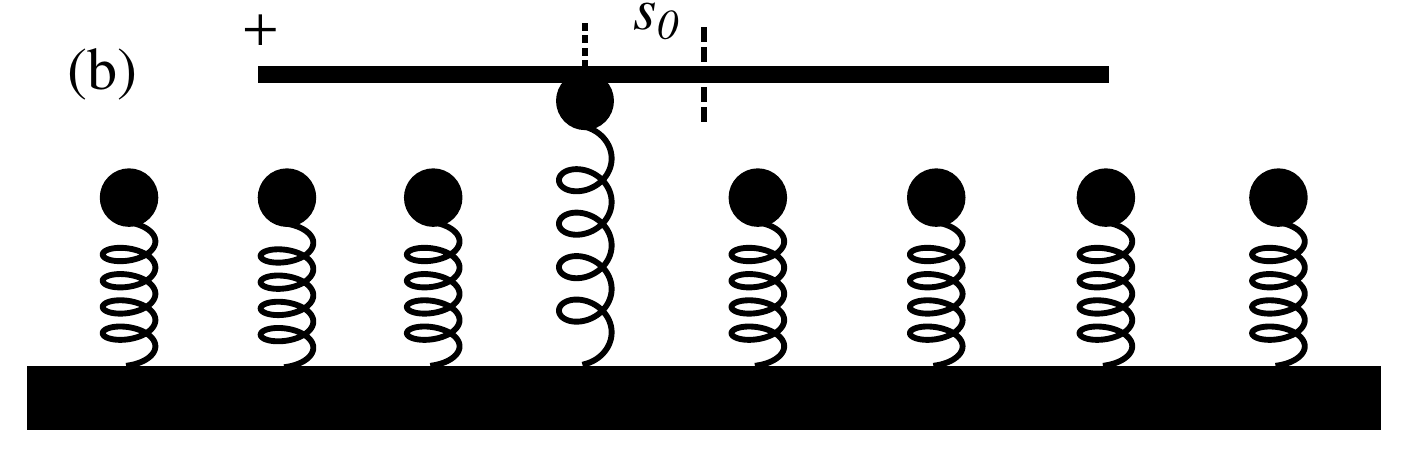}\vspace{0.2in}\\
\includegraphics[width=7cm]{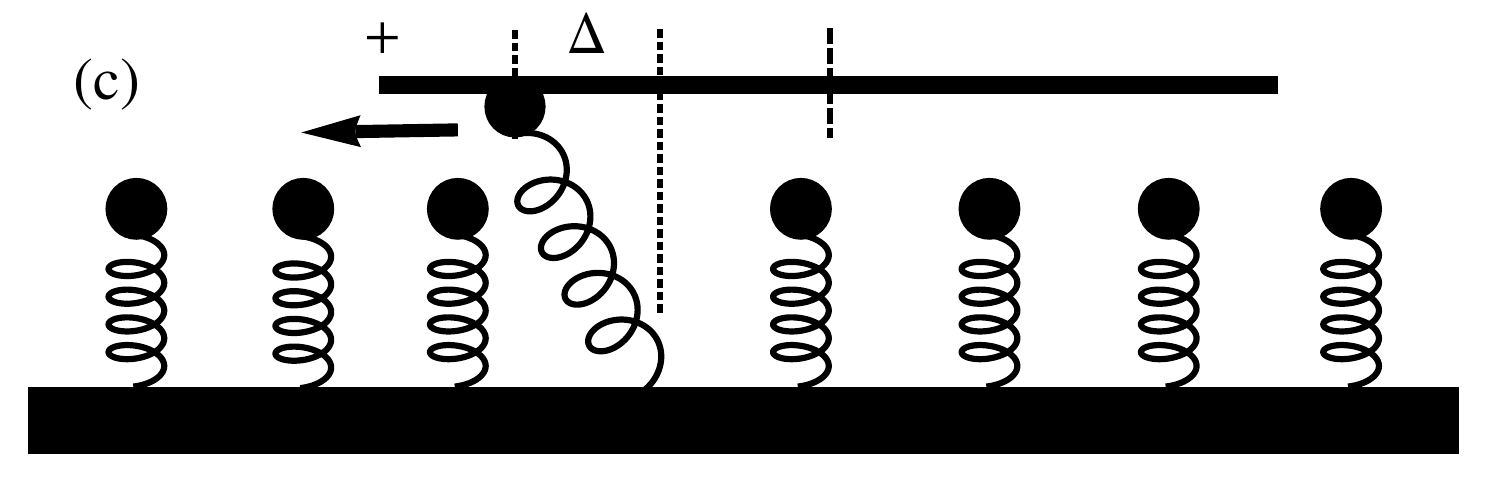}\vspace{0.2in}\\
\includegraphics[width=7cm]{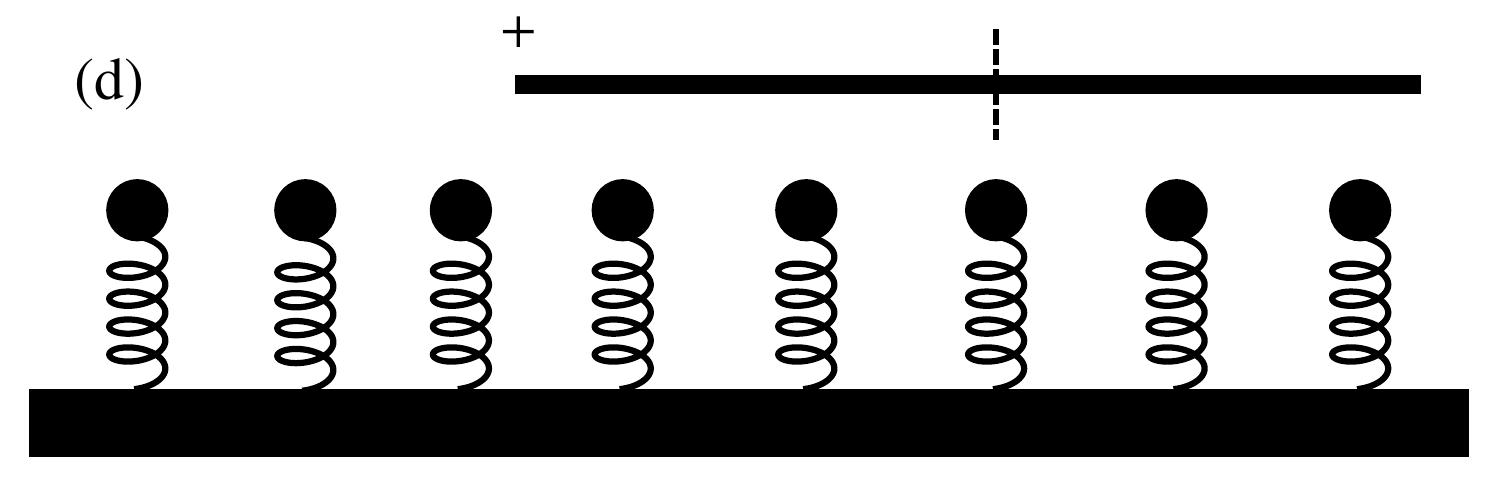}
\end{center}
\caption{The figure shows the four steps of a motor cycle. In (a) a filament is
sliding with velocity $v$ over a uniform density of unbound motors with tails
tethered to the substrate. In (b) a motor attaches to the filament at a
position $s_0$ from the filament's mid-point. The stretch of the motor tails at
the time of attachment is neglected.  In (c) the motor has walked towards the
polar head of the filament, stretching the tails by an amount $\Delta$.
Finally, in (d) the bound motor detaches and relaxes instantaneously to its
unstretched state. The filament has undergone a net displacement in the
direction opposite to that of motor motion. 
} \label{motor-cycle}
\end{figure}

We model the interaction cycle
of an individual motor protein with the filament as shown in
Fig.~\ref{motor-cycle} for a one-dimensional system.   
The tail of a specific motor is fixed at position ${\bf x}_{t}$ in the plane.
At a time $t_{0}$ the head of this motor attaches to a point on the filament.
The position of the motor
head at the time of attachment  is ${\bf x}_{h}(t_{0})={\bf r}(t_{0})+s_{0}{\bf
\hat{u}}(t_{0})$, where ${\bf
r}(t_0)$ and ${\bf \hat{u}}(t_0)$ denote the position of the center of the filament  and its orientation  $t=t_0$ and $s_{0}\in \left[-L/2,L/2\right]$ parametrizes the
distance of the point of attachment from the center  of the filament 
(cf. Fig.~\ref{motor-cycle}(b)).  We
assume that 
motor proteins will attach to parts of the filament which are within a distance
of the order of the size of the motor protein.
The stretch of the motor tail at the time of attachment is then of order of the
motor size and will be neglected, \emph{i.e.} ${\bf x}_h(t_0)-{\bf x}_t=0$, or
motors attach to the part of the filament directly overhead without any initial
stretch.

\begin{figure} 
\begin{center}
\includegraphics[scale=0.8]{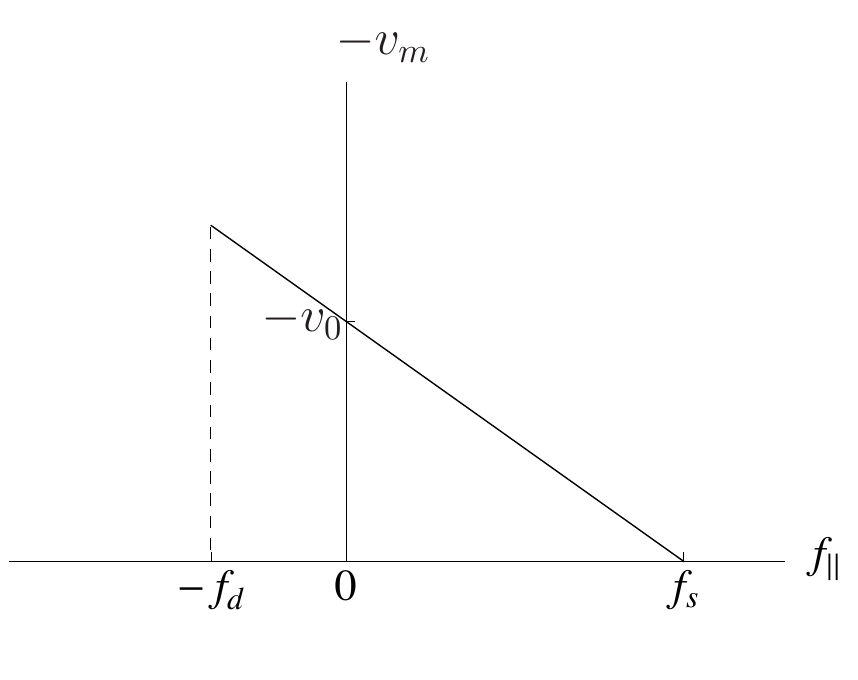} 
\end{center}
\caption{The velocity $-v_m$ of a loaded motor head as a function of the load
$f_\Vert={\bf \hat{u}}\cdot\bm\Delta$. The figure shows the stall force $f_s$
where $v_m=0$ and the detachment force $-f_d$.}
\label{Fig:vm} 
\end{figure}

For $t>t_{0}$ the motor head remains attached to the filament and walks along
it towards the polar head ($-{\bf \hat{u}}$ direction) until detachment.  The
tails, modeled as a linear spring of force constant $k$, exert a load ${\bf
f}=-k\bm\Delta(t,\tau;s_0)$ on the head, where $\bm\Delta(t,\tau ;s_0)={\bf
x}_{h}(t)-{\bf x}_{t}$ is the stretch at time $t$ of a motor protein that has
been attached for a time $\tau$, \emph{i.e.} $t=t_0+\tau$
 (cf. Fig.~\ref{motor-cycle}(c)).  Since we assume
$\bm\Delta(t_{0})=0$, we can also write
\begin{eqnarray} 
\label{Delta} 
\bm\Delta(t,\tau;s_0)&=&{\bf r}(t)-{\bf
r}(t-\tau)+\sigma(t,\tau){\bf \hat{u}}(t)\nonumber\\ &&+s_0\left[{\bf
\hat{u}}(t)-{\bf \hat{u}}(t-\tau)\right]\;, 
\end{eqnarray}
where $\sigma(t,\tau)=s(t)-s(t-\tau)$ is the distance traveled along the
filament at time $t$ by a motor head that has been attached for a time $\tau$,
measured from the initial attachment position, $s_0$.  The kinematic constraint
imposed by the condition of attachment requires
\begin{eqnarray} 
\label{dot-Delta} 
\partial_t{\bm\Delta}(t,\tau;s_0)&&={\bf
v}(t) -{\bf v}(t-\tau) +{\bf \hat{u}}(t)\left[v_m(t)-v_m(t-\tau)\right]
\nonumber\\ &&+
\bm\Omega(t)\sigma(t,\tau)+s_{0}\left[\bm\Omega(t)-\bm\Omega(t-\tau)\right]\;,
\end{eqnarray}
where $\bm\Omega(t)=\partial_t{\bf \hat{u}}(t)=\dot{\theta}{\bf \hat{n}}(t)$ is
the angular velocity of the rod and $v_m(t)=\partial_t s(t)$ the velocity of the
motor head along the filament. We have introduced a unit vector ${\bf
\hat{n}}={\bf \hat{z}}\times{\bf \hat{u}}$ normal to the long axis of the
filament. Then $({\bf \hat{z}},{\bf \hat{u}},{\bf \hat{n}})$ defines a
right-handed coordinate system with in-plane axes longitudinal and transverse
to the filament. We note that Eq.~\eqref{dot-Delta} can also be written as
\begin{eqnarray} 
\label{dot-Delta-2}
\partial_t{\bm\Delta}(t,\tau;s_0)&+&\partial_\tau{\bm\Delta}(t,\tau;s_0)={\bf
v}(t)  +v_m(t){\bf \hat{u}}(t) \nonumber\\ &&+\bm\Omega(t)\sigma(t,\tau)+
s_{0}\bm\Omega(t)\;.  
\end{eqnarray}
While the motor remains bound, the dynamics of the motor head along the
filament is described by an overdamped equation of motion
\begin{equation} 
\label{vm} 
\zeta_m\dot{s}(t)=-f_s+{\bf \hat{u}}\cdot{\bf f}
\end{equation}
where $f_s>0$ is the stall force, defined as the force where the velocity
$v_m=\dot{s}$ of the loaded motor vanishes.  Since motors move in the
$-{\bf \hat{u}}$ direction, generally $v_m=\dot{s}<0$.  Letting
$f_{\Vert}={\bf \hat{u}}\cdot{\bf f}=-k\Delta_\Vert$, Eq.~\eqref{vm} can also
be written as
\begin{equation} 
\label{eq:vm}
v_m(t)=-v_0\left(1-\frac{f_\Vert(\Delta_\Vert)}{f_s}\right)\;, 
\end{equation}
where  $v_0=f_s/\zeta_m\sim\Delta\mu>0$ is the load-free stepping velocity,
with $\Delta\mu$ the rate of ATP consumption. The motor velocity is shown in
Fig.~\eqref{Fig:vm} as a function of the load $f_\Vert$. The motor head
velocity also vanishes for $f_\Vert<-f_d$, when the motor detaches. The linear force-velocity relation for an individual motor is consistent with experiments on single kinesin molecules~\cite{Svoboda1994}.

The active force and torque on the filament due to an individual bound motor
can then be expressed in terms of these quantities as
\begin{subequations} 
\begin{gather}
{\bf f}_a(t,\tau;s_0)=-k\bm\Delta(t,\tau;s_0)\;,\\
\tau_a(t,\tau;s_0)=-{\bf \hat{z}}\cdot\left[(s_0+\sigma(t,\tau)){\bf
\hat{u}}(t)\times k\bm\Delta(t,\tau;s_0)\right]\;.
\end{gather}  
\end{subequations}
Finally, after traveling along the filament for a time $\tau_\text{detach}$, the
motor head detaches and the head position relaxes instantaneously back to the
fixed position ${\bf x}_{t}$ of the tail.

We note that we shall not be considering the possibility of direct interactions
of motors with each other.  We have also not considered stochastic aspects of
the motor motion along the filament (Eq.~\eqref{vm}). 

\subsection{Motor binding and unbinding} 
\label{Sec:BindingUnbinding}

Next we need to describe the stochastic binding/unbinding dynamics of the motor
heads.  We assume the motor tails are attached to the substrate with a
homogeneous surface density $\rho_m$, such that for  a rod of length $L$ and
width $b$ a maximum of $N=\rho_m Lb$ motors can be bound at any given time.  Following
Gu\'erin et al.~\cite{Guerin2010}, we denote by ${\cal P}_b(t,\tau;s_0)$ the
probability that a motor head that has attached at $s_0$ at a time $t_0$, has remained attached
for a duration $\tau$ at time $t$. For simplicity in the
following we assume that the probability that a motor attaches at any point
along the filament is uniform, i.e., ${\cal
P}_b(t,\tau;s_0)=\frac{1}{L}P_b(t,\tau)$.  We further assume that when motors
unbind they relax instantaneously to the unstretched state.  The time evolution
of the binding probability is then given by
\begin{align}
\label{P-u} 
\partial_t{ P}_b(t,\tau)+\partial_\tau {
P}_b(t,\tau)=&-\langle\omega_u(\bm\Delta(\tau))\rangle_{s_0}{P}_b(t,\tau)\notag\\
&+
\omega_b\delta(\tau)p_u(t)\;,
\end{align}
where $p_u(t)$ is the probability that a motor be unbound at time $t$.  The
probability distribution is normalized according to
\begin{equation} \label{P-norm}
\int_{0}^{\infty}d\tau\int_{-L/2}^{L/2}ds_0~{\cal P}_b( t,\tau;s_0)+p_u(
t)=1\;.  
\end{equation}
In Eq.~\eqref{P-u}, $\omega_u(\bm\Delta(\tau))$ and $\omega_b$ are the rates at
which a motor head with tails stretched by an amount $\bm\Delta(t,\tau)$
unbinds from and binds to the filament, respectively. The binding rate
$\omega_b$ will be assumed to be constant. In contrast, the unbinding rate
$\omega_u$ is a strong function of the stretch of the motor tails, that has to
be obtained by solving Eq.~\eqref{dot-Delta-2}, with initial condition
$\Delta(t=0,\tau)=0$.  We will see below that the nonlinear dependence of the
unbinding rate plays an important role in controlling the filament dynamics. In
two dimensions the unbinding rate $\omega_u$ also depends on the initial
attachment point $s_0$ along the filament. To be consistent with our ansatz
that the probability that the motor attaches at any point along the filament is
uniform, we have replaced the rate in Eq.~\eqref{P-u} with its mean value
$\langle\omega_u\rangle_{s_0}$, where $\langle
...\rangle_{s_0}=\int_{-L/2}^{L/2}\frac{ds}{L}...$ denotes an average over the
initial attachment points.  

The unbinding rate is controlled by the work done by the force (load) acting on
the motor head, which in turn is a linear function of the stretch $\bm\Delta$.
A form that has been used extensively in the literature for one-dimensinal
models is an exponential, $\omega_u=\omega_0e^{\alpha|\Delta|}$,  where
$\omega_{0}$ is the unbinding rate of an unloaded motor and  $\alpha$ is a
characteristic length scale that control the maximum stretch of the tails above
which the motor unbinds~\footnote{$\alpha$ can be estimated to be equal to $ka / k_B T$, where $a$ is a microscopic length scale of the order of a few nm. Experiments are carried out at room temperatures which leads to $k_B T \sim pN nm$.}.  The exponential form represents an approximation for
the result of a detailed calculation of the average time that a motor moving
along a polar filament spends attached to the filament as a function of a
tangentially applied load~\cite{Parmeggiani2001} and is consistent with
experiments on kinesin~\cite{Visscher1999}. This form can easily be generalized to to
the case of a filament sliding in two dimensions where the motor load had both
components  tangential and transverse to the filament. It
is, however, shown in the Appendix that within the mean-field approximation used
below the exponential form yields a steady-state  stretch $\Delta$ that
saturates to a finite value at large velocity $v$ of the filament. This is
unphysical as it does not incorporate the cutoff described by the detachment
force $f_d$ in Fig.~\ref{Fig:vm}. For this reason in the mean-field treatment
described below we use a parabolic form for the unbinding rate as a function of stretch,
\begin{equation}
\omega_u(\bm\Delta)=\omega_0\left[1+\alpha^2|\bm\Delta|^2\right]\;,
\label{omegau-assumption-par} \end{equation}
where for simplicity we have assumed an isotropic dependence on the magnitude
of the stretch in terms of a single length scale, $\alpha^{-1}$. An explicit comparison of the two expressions for the unbinding rates is
given in the Appendix.

The total active force and torque on the filament averaged over
the original positions and the times of attachment can be written as
\begin{widetext} 
\begin{subequations} 
\begin{gather} 
{\bf F}_a(t)=
-N k\int_{0}^{\infty}d\tau~ \langle P_b(t,\tau)~
\bm\Delta(t,\tau;s_0)\rangle_{s_0}\;,
\label{Fa}\\ 
T_a(t)=
-N k\int_{0}^{\infty}d\tau ~\langle P_b(t,\tau)~{\bf
\hat{z}}\cdot\left[(s_0+\sigma(t,\tau)){\bf \hat{u}}(t)\times
\bm\Delta(t,\tau;s_0)\right]\rangle_{s_0}\;.
\label{taua} 
\end{gather}
\end{subequations}
 \end{widetext}

\section{Mean field approximation} 
\label{Sec:MeanField}

To proceed, we introduce several approximations for the motor dynamics.  First,
we restrict ourselves to the dynamics on times scales large compared to the
attachment time $\tau$ of individual motors. For $t\gg\tau$ we
approximate
\begin{subequations} 
\begin{gather} 
\sigma(t,\tau)\simeq
v_m(t)\tau\;,
\label{sigma-app}\\ 
\bm\Delta(t,\tau;s_0)\simeq\left[{\bf
v}(t)+v_m(t){\bf \hat{u}}(t)+s_0\bm\Omega(t)\right]\tau\;.
\label{Delta-app}
\end{gather} 
\end{subequations}
This approximation becomes exact for steady states where the filament and
motor velocities are independent of time.  We also stress that in
Eqs.~\eqref{sigma-app} and \eqref{Delta-app}  $\sigma$ and $\bm\Delta$ are
still nonlinear functions of $\tau$ due to the dependence of $v_m$ on the load
force. 

Secondly, we recall that we have assumed that the attachment positions $s_{0}$
are uniformly distributed along the filament and  can be treated as independent
of the residence times $\tau$.  Finally, we make a mean field assumption on the
probability distribution of attachment times, which is chosen of the form
$P(t,\tau)=\delta(\tau-\tau_\text{MF})p_b(t)$, with $p_b(t)$ the probability that a
motor be attached at time $t$ regardless of the its attachment time. The
mean-field value of the attachment time is determined by requiring 
\begin{equation} 
\label{tauMF}
\tau_\text{MF}=\left[\langle\omega_u
\left(\Delta(\tau_\text{MF})\right)\rangle_{s_0}\right]^{-1}\;.
\end{equation}
In previous literature a similar mean field assumption has been stated in terms
of the stretch, $\bm\Delta$~\cite{Grill2005,GuntherKruse2007}. In the present problem, however, where filaments
can slide in two dimensions, it is necessary to restate the mean-field theory
in terms of the residence time $\tau$ as the active forces and torques depend
on both the stretch $\bm\Delta$ of the motor tails and the distance $\sigma$
traveled by a bound motor head along the filament. These two quantities are in
turn both controlled by a single stochastic variable, identified with the
residence time $\tau$.  The rate of change of the probability $p_b(t)$ that a
motor be bound at time $t$ is then described by the equation
\begin{equation} 
\label{p-b} 
\partial_t{
p}_b(t)=-\tau_\text{MF}^{-1}p_b(t)+\omega_b\left[1-p_b(t)\right]\;, 
\end{equation}

The mean field active  force and torque due to the motors are then given by
\begin{widetext}
\begin{eqnarray} 
\label{forceMF} 
&& {\bf F}_a^\text{MF}= -k N  \langle
\bm\Delta(t,\tau_\text{MF};s_0) p_b(t)\rangle_{s_0}\;,
\\
 &&T_a^\text{MF}=-k N \langle
p_b(t)~{\bf \hat{z}}\cdot\left[(s_0+\sigma(t,\tau_\text{MF})){\bf \hat{u}}(t)\times
\bm\Delta(t,\tau_\text{MF};s_0)\right]\rangle_{s_0}\;.\label{torqueMF}
\end{eqnarray}
\end{widetext} In the following we will work in the mean-field approximation
and remove the label MF from the various quantities.


\section{Active Filament Sliding in One Dimension} 
\label{Sec:ActiveFilament 1D}

We first consider the
simplest theoretical realization of a motility assay experiment, where the
actin filament is sliding over a one dimensional track of tethered motor
proteins. A closely related model, where the filament is elastically coupled to
a network, has been used extensively in the literature to study the onset of
spontaneous oscillations arising from the collective action of the bound
motors~\cite{JulicherProst1995,JulicherProst1997,Grill2005}. Previous studies
of freely sliding filaments, as appropriate for the modeling of motility
assays, have also been carried out both analytically and
numerically~\cite{Kafri2009}. Our work contains some new results on the
response to an external force of a filament free to slide under the action of active
crosslinkers and also on the filament fluctuations.

The
Langevin equation for the center of mass coordinate $x$ of the filament is
given by
\begin{equation} 
\label{xdot} 
	\zeta \dot{x} = F_\text{a}(t) + F_\text{ext} + \eta(t)\;,
\end{equation}
where $\dot{x}$ is the center-of-mass velocity of the filament and the
mean-field active force is given by
\begin{equation} 
	F_{a}^\text{MF}(t)=-kNp_b(t)\Delta(\dot{x},\tau)\;.
\end{equation}
In one dimension the dependence on $s_0$  drops out and Eq.~\eqref{Delta-app}
simply gives $\Delta\simeq (\dot{x}+v_m)\tau$. Substituting Eq.~\eqref{eq:vm} for
$v_m$, we can solve for $\Delta$ as a function of $\dot{x}$ and $\tau$,
\begin{equation} 
\label{Delta-1d} 
	\Delta(\dot{x},\tau)=\frac{(\dot{x}-v_0)/\omega_0}{\tilde{\tau}^{-1}+\epsilon}\;, 
\end{equation}
and Eq.~(\ref{tauMF}) for the mean attachment time becomes
\begin{equation}
	\tilde\tau^{-1}(\dot{x})=1+\frac{(\dot{x}-v_0)^2\alpha^2}{\left[\tilde\tau^{-1}(\dot{x})+
	\epsilon\right]^2\omega_0^2}\;,
\label{tauMF1d} 
\end{equation}
where $\tilde\tau=\omega_0\tau$ and
$\epsilon=kv_0/f_s\omega_0$.  The parameter $\epsilon$ is the ratio of the length $\ell_0=v_0/\omega_0$ traveled by an unloaded
motor that remains attached for a time $\omega_0^{-1}$ to the stretch $\delta_s=f_s/k$ of the motor tails at the stall force, $f_s$. 
Typical values for these length scales and the parameter $\epsilon$ are given in Table~\ref{parameters-motor}. 
\begin{table}[tp]
\centering
\begin{tabular}{| c | c | c |}
\hline
Parameters & Myosin-II & Kinesin\\
\hline
$\ell_0$ &   $\sim 2 \text{ nm}$  &  $\sim 8 \text{ nm}$  \\
\hline
$\delta_s$ &  $ \sim 1 \text{ nm}$ & $\sim 25 \text{ nm}$ \\
\hline
$\epsilon$ &  $ \sim 2$  &   $\sim 0.32$      \\
\hline
\end{tabular}
\caption{Typical values of the length scales $\ell_0=v_0/\omega_0$ and $\delta_s=f_s/k$ introduced in the text and the ratio $\epsilon$ for myosin II and kinesin. The parameters are taken from Refs.~\cite{Howard2001} and~\cite{Gibbons2001}.}
\label{parameters-motor}
\end{table}

It is convenient to rewrite the mean residence time $\tilde\tau$ as
\begin{equation}
\label{tautilde}
\tilde\tau^{-1}=1+\frac{(u-1)^2\nu^2}{\left[\tilde\tau^{-1}+
	\epsilon\right]^2}\;,
	\end{equation}
where $u=\dot{x}/v_0$	 and we have introduced a dimensionless parameter $\nu=\ell\alpha$ that controls the dependence of the unbinding rate on the load exerted on the bound heads by the stretched motor tails, with 
\begin{equation}
\frac{1}{\ell}=\frac{1}{\ell_0}+\frac{1}{\delta_s}
\label{ell}
\end{equation}
  the geometric mean of the two length scales introduced earlier. For stiff motors, with $\epsilon \gg1$ or  $\ell_0 \gg \delta_s$, $\ell\sim \delta_s$, while for floppy, easy to stretch motors, corresponding to 
	$\epsilon \ll 1$ or  $\ell_0 \ll \delta_s$, $\ell\sim \ell_0$.
	Setting $\nu=0$ corresponds to neglecting the load dependence of the unbinding rate. 
The exact solution to Eq.~\eqref{tautilde}  for
the mean residence time $\tilde\tau(\dot{x})$ as a function of the filament velocity 
can be determined and is discussed in the Appendix.
Clearly $\tau$ has a maximum
value at $\dot{x}=v_0$, where $\tau=\omega_0^{-1}$ and decays rapidly as $|\dot{x}-v_0|$
grows.

\subsection{Steady State and its Stability} 
\label{Sec:SteadyState1D}

We begin by characterizing the steady state dynamics of the filament  in the
absence of noise.  Incorporating for generality an external force
$F_\text{ext}$, the steady state velocity  $v$  of the filament is obtained
from the solution of the nonlinear equation 
\begin{equation} 
	\zeta v=F_\text{ext}+F_a(v)
	\label{steady-state}
\end{equation}
where $F_a(v)=-kNp_{bs}(v)\Delta(v)$. The steady state stretch $\Delta(v)$ is 
 given by Eq.~\eqref{Delta-1d} with $\dot{x}=v$ and 
\begin{equation} 
	p_{bs}(v)=\frac{ \omega_b\tau(v)}{1+\tau(v)\omega_b}\;,  
\label{Nbs}
\end{equation}
with $\tau(v)$ given by Eq.~\eqref{tautilde} for $\dot{x}=v$.
To gain some insight in the behavior of the system, we expand the active force
as $F_a(v)\simeq F_{p}+\left(\frac{\partial F_a}{\partial v}\right)_{v=0}v +
{\cal O}(v^2)$, with $F_{p}=F_a(v=0)$. Retaining only terms linear in $v$ this
gives a steady state force/velocity relation of the form
\begin{eqnarray} (\zeta+\zeta_a)v=F_\text{ext}+F_{p}
\label{definition-linear-friction} \end{eqnarray}
with a filament ``propulsion'' force $F_{p}$
\begin{equation} 
\label{Fst}
F_{p}=\frac{Np_{bs0} k\ell_0}{
\epsilon+\tilde\tau_0^{-1}}\;,
\end{equation}
where $p_{bs0}=r/[r+(1-r)\tilde\tau_0^{-1}]$, with $r=\omega_b/(\omega_0+\omega_b)$  the duty ratio, and $\tilde\tau_0=\tilde\tau(v=0)$. The active
contribution $\zeta_a= -\left(\frac{\partial
F_a}{\partial v}\right)_{v=0}$ to the friction is given by
\begin{eqnarray} 
\label{zeta-a} 
\zeta_a=Np_{bs0}\frac{k|\Delta_0|}
{v_0}\left[1-\left(\frac{|\Delta_0|}{\ell_0}+p_{bs0}\frac{1-r}{r}\right)\frac{2\alpha^2\Delta_0^2\ell_0}
{\ell_0+2\alpha^2|\Delta_0|^3}\right]\;,
\end{eqnarray}
where $\Delta_0=\Delta(v=0)=-\ell_0/(\tilde\tau_0^{-1}+\epsilon)$. In the absence of external force, the filament will slide
at a velocity 
\begin{equation}
\label{vs}
v_{s}=F_p/(\zeta+\zeta_a)
\end{equation}
 due to the action of the motor
proteins. This motion is in the polar direction of the filament and opposite to
the direction of motion of bound motors along the filament. Phenomenological
models of motility assays have described the actin filaments as
``self-propelled'' Brownian rods. Our model yields a microscopic calculation of
such a ``self-propulsion'' force $F_p$ in terms of microscopic parameters
characterizing  the motor proteins. We note that $-F_p$ can also be interpreted
as the ``stall force'' of the filament, \emph{i.e.} the value of $F_\text{ext}$ 
required to
yield $v=0$. This is a quantity that may be experimentally accessible using
optical force microscopy.

If we neglect the load dependence of the unbinding rate by letting $\nu=0$, the mean number of bound motors is simply $Nr$  and $F_{p}^{0}=Nrk \ell$, with $\ell$ given by Eq.~\eqref{ell}.  In this limit the sliding
velocity $v_{s}^{0}$ in the absence of external force can be written as
\begin{equation} \label{vs0}
v_{s}^{0}=\frac{v_0}{1+\zeta/\zeta_{a}^{0}}\;.
\end{equation}
where the active friction $\zeta_{a}^{0}=Nrk\ell/v_0>0$ is always positive.
The sliding velocity vanishes when $v_0\rightarrow 0$ and it saturates to its
largest value $v_0$ when the number $N r$ of bound motors becomes very large and
$\zeta_{a}^{0} \gg \zeta$. The behavior is controlled by the parameter $\epsilon$. If
the motors are easy to stretch, i.e.,  $\epsilon \ll 1$, then  the propulsion
force is determined entirely by the elastic forces exerted by these weak bound
motors, with $F_p^0\simeq N rk\ell_0$. 
On the other hand stiff
motors, with $\epsilon \gg 1$,  stall before detaching. The propulsion force is
then  controlled by the motor stall force, with $F_\text{p}^0\simeq N rf_s$.

\begin{figure} 
\begin{center}
 \includegraphics[scale=0.65]{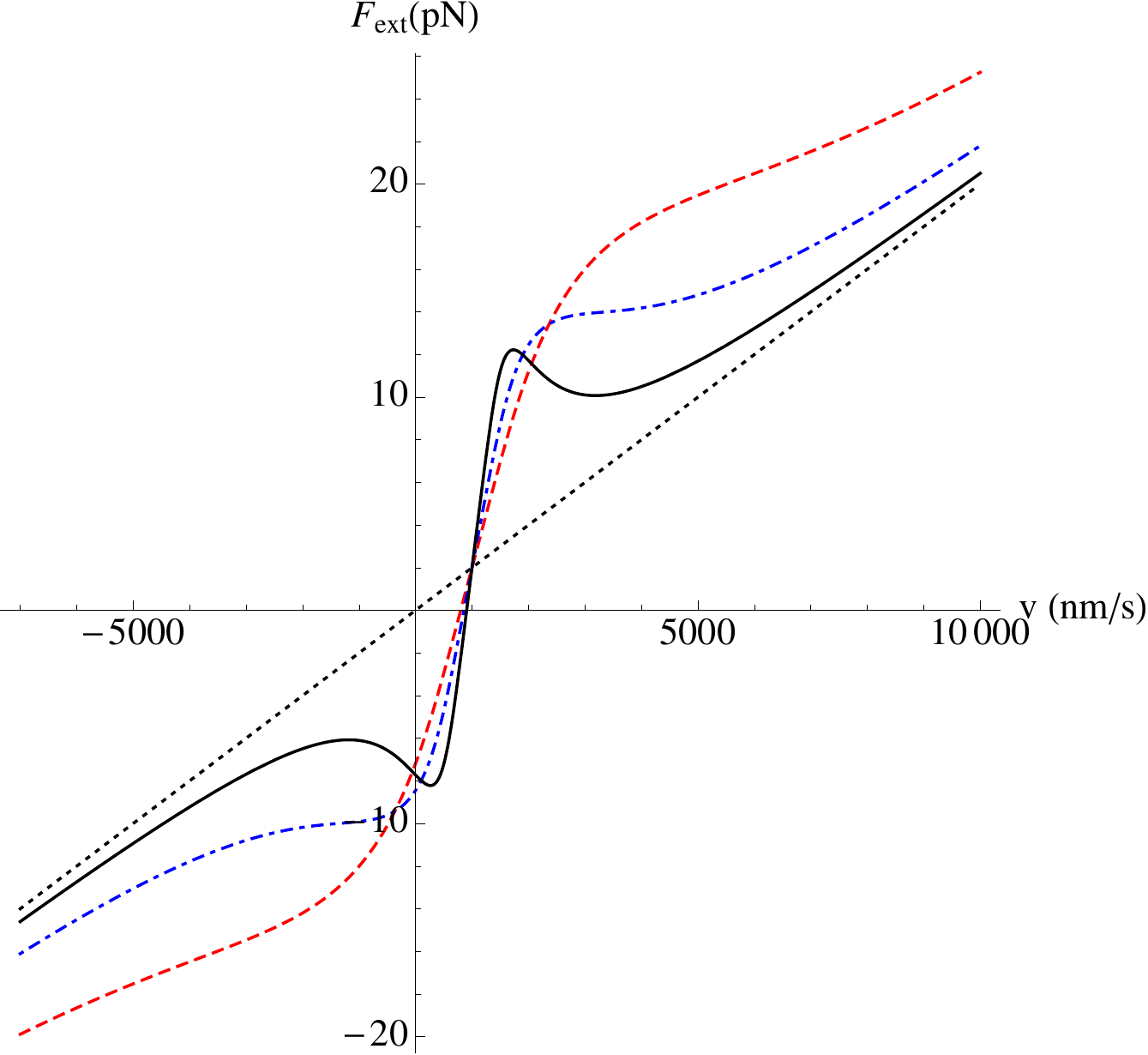}
\end{center} 
\caption{(Color online) Force-velocity curves for $\zeta=0.002\ \text{pN}  \text{nm}^{-1}\text{s}$ and various values of the motor
stiffness $k$, showing the
transition to non-monotonicity as $k$ increases. The values of the stiffness $k$ (in pN/nm) and the corresponding values for $\alpha^{-1}$ (in nm) and $\epsilon$ are as follows: 
$k=0$, $\alpha^{-1}=0$, $\epsilon=0$ (black dotted line); $k=1$ , $\alpha^{-1}=0.75$, $\epsilon=0.5$ (red dashed line);
$k=2$, $\alpha^{-1}=1.5$, $\epsilon=1$ (blue dashed-dotted line);
$k=8$, $\alpha^{-1}=6$, $\epsilon=4$ (black solid line).
At high velocities the curves merge into the linear curve 
$F_\text{ext}=\zeta v$ (black dotted line), corresponding to the case where no motors are present.  
The remaining parameters have the following values:  $N=\rho_m Lb=100$, $v_0=1000\ \text{nm/s}$, $f_s=4\
\text{pN}$, $\omega_0=0.5\ (\text{ms})^{-1}$, $r=0.06$.}
\label{forcevelocity} 
\end{figure}

\begin{figure} 
\begin{center} 
\includegraphics[scale=0.53]{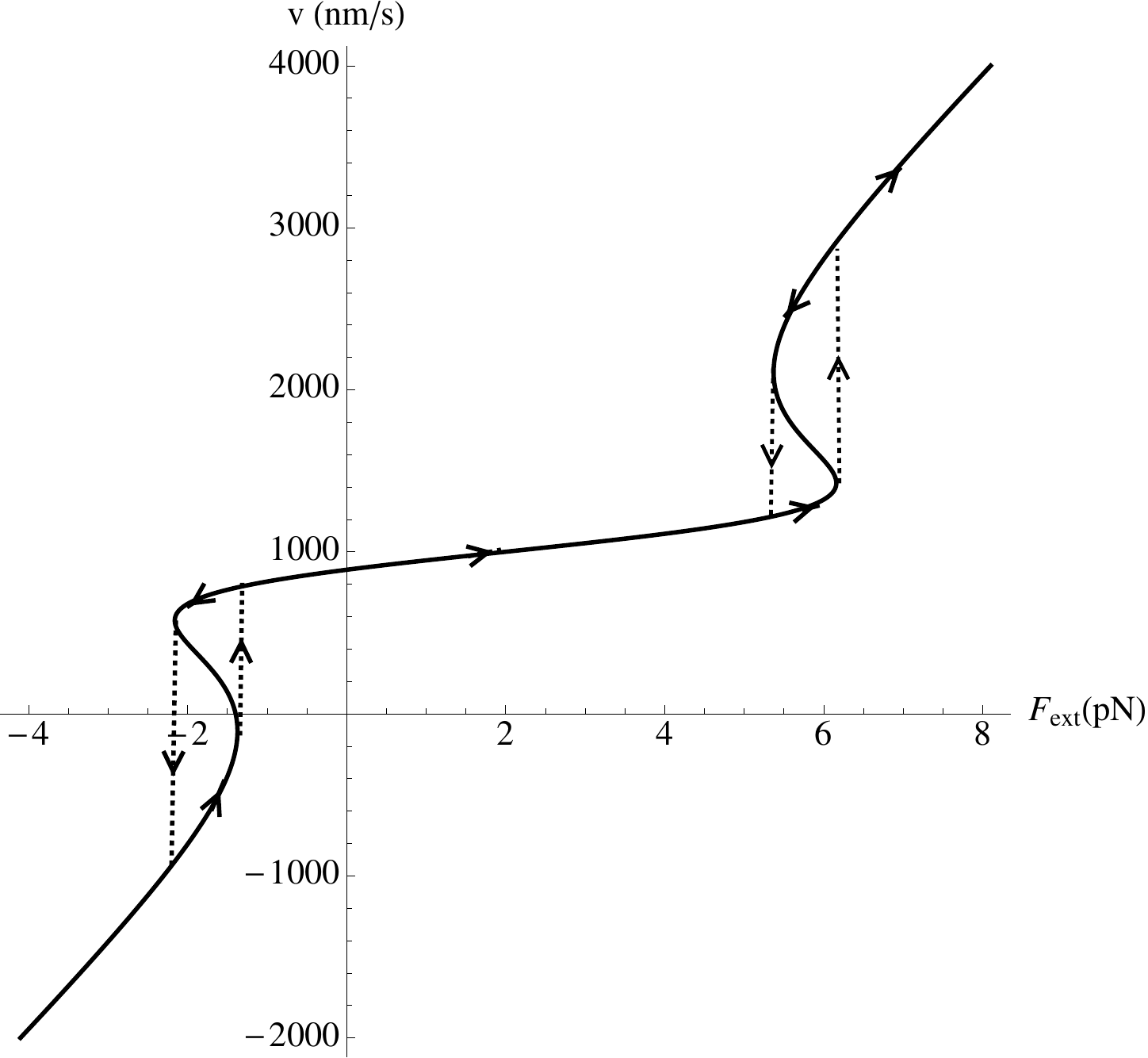}
\end{center} 
\caption{The figure sketches the hysteretic behavior that may be
obtained in an experiment where an external force $F_\text{ext}$ is applied to a
filament in a motility assay.  The response of the filament will generally
display two regions of hysteresis, at positive and negative forces. }
\label{hysteresis} 
\end{figure}

The load-dependence of the unbinding rate changes qualitatively the behavior of
the system. In particular, the net friction $\zeta+\zeta_a$ can become
negative, rendering the steady state unstable. This instability was already
noted in Ref.~\cite{JulicherProst1995} for a two-state model of active linkers
and in Ref.~\cite{Guerin2010} for a two state ``soft'' motor model. The full
nonlinear force-velocity curves are shown in Fig.~\ref{forcevelocity} for
various values of the motor stiffness $k$, for parameters appropriate for
acto-myosin systems. In the steady state, as we increase the active parameter
$k$ while keeping the substrate friction $\zeta$ constant, the $F_\text{ext}-v$
curve becomes non-monotonic, and two distinct regions of bistability emerge. 
To understand the increase of the bistability region with motor stiffness, we note that the active force is simply proportional to $k$, hence naively one would indeed expect its effect to be more pronounced for stiff motors. The detailed behavior is, however, determined by the interplay of the mean residence time $\tau$ that motors spend bound to the filament and the stretch, $\Delta$.  Soft, floppy motors have large stretches, controlled mainly be the length $\ell_0$ traveled by an unloaded motors advancing at speed $v_0$. On the other hand, their residence time is small  and the overall effect of the active force remains small. In contrast, stiff motors have a small stretch, of order of the  stretch $\delta_s=f_s/k$  of a stalled motor, but long residence times and are collectively capable of slowing down the filament and even holding it in place against the action of the external force, driving the negative friction instability. At even larger values of the external force motors are effectively always unbound  due to the fast sliding of the filament and the velocity-force curve approaches the linear form obtained when no motors are present. This behavior is best seen from Fig.~\ref{hysteresis}.

\begin{figure} 
\begin{center} 
\includegraphics[scale=0.9]{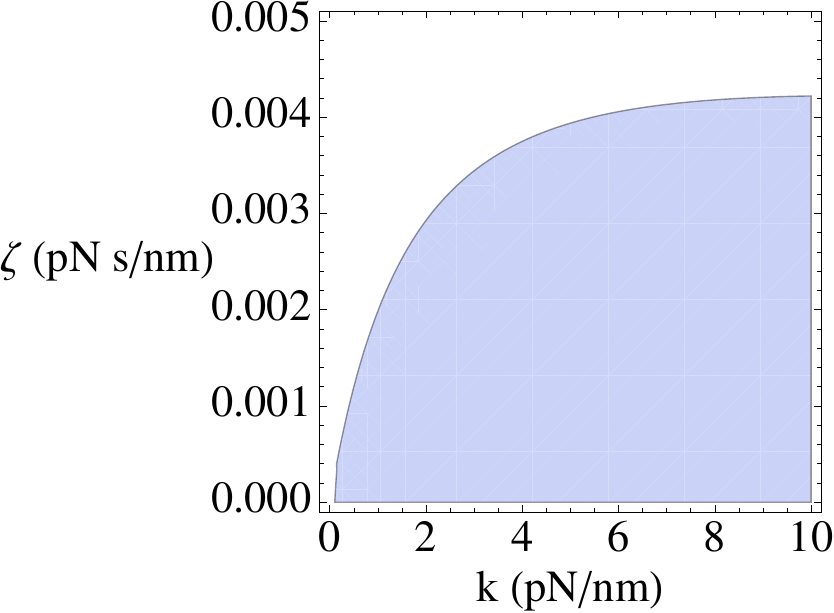}
\end{center} 
\caption{(Color online) "Phase diagram" in $k$-$\zeta$ plane showing the region where the $F_\text{ext}$-$v$ curves exhibit non-monotonic
behavior  (blue shaded region)  for  $N=\rho_m Lb=100$ and $v_0=1\ \mu\text{m s}^{-1}$, $f_s=4\ \text{pN}$, 
$\alpha/k=1.33\ \text{pN}$, $\omega_0=0.5\ (\text{ms})^{-1}$, $r=0.06$.}
\label{phase} 
\end{figure}

The region of non-monotonicity of the force-velocity curve and associated
bistability can also be displayed as a phase diagram, as shown in
Fig.~\ref{phase}.  The stiffness of myosins is about $5$ pN/nm and the actin filament friction was estimated to be of order $0.003$ pNs/nm in Ref~\cite{Riveline1998}. In actomyosin systems the negative friction instability should therefore be observable in a range of experimentally relevant parameters. Kinesin motors have floppier tails and a smaller stiffness of about $0.5$ pN/nm. In this case bistability effects should be prevalent only at very low filament friction, $\zeta\ll 0.001$ pNs/nm. A proper estimate of the region of parameters where the instability may be observable is rendered difficult by the fact that the onset of negative friction is also a strong function of the  density of motors tethered to the substrate, which in turn affects the value of the friction $\zeta$. In general, we expect that a high motor density will be needed for the instability to occur. On the other hand, if the density of motors is too high, the friction $\zeta$ will be enhanced and the instability suppressed.

We stress that the force-velocity curves displayed in Fig.~\ref{forcevelocity}
have been obtained by calculating $F_\text{ext}$ as a function of $v$. In an
experiment one would tune the applied force and measure the resulting velocity.
The system would not access the unstable regions of negative friction, but
rather follow the hysteretic path sketched in Fig.~\ref{hysteresis}. The
discontinuous jump may occur at the boundary of the stability region, as shown
in the figure, or before such a boundary is reached, corresponding to what is
known as  ``early switching''.

\begin{figure} 
\begin{center} 
\includegraphics[scale=0.75]{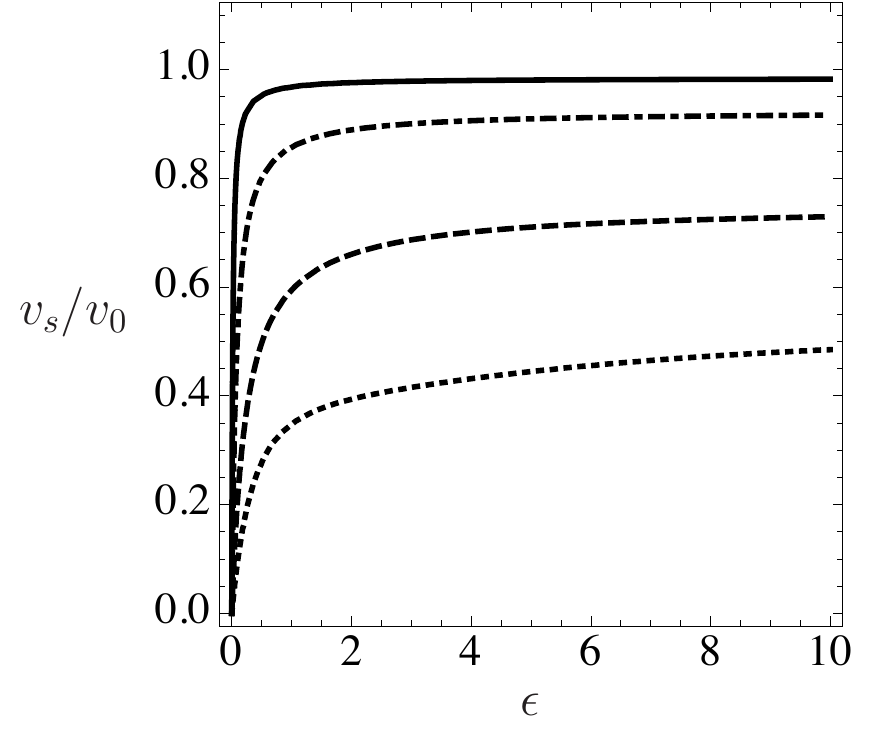}
\end{center}
 \caption{The motor-induced sliding velocity $v_s$ of an actin
filament in the absence of external force is shown as a function of
$\epsilon=\ell_0/\delta_s$ for $N=10$ (dotted line),  $N=25$ (dashed line),
$N=100$ (dashed-dotted line) and $N=500$ (solid line). We observe that
$v_s\rightarrow v_0$ for stiff motors as $N$ is increased. Parameter values: $\zeta=0.002\ \text{pN }(nm)^{-1}\text{s}$, $r=0.06$, $\alpha/k=1.33\ \text{pN}$. } 
\label{fig:sliding-vel}
\end{figure}

To summarize, motors have two important effects on the steady state dynamics of
the filament. First, they make the filament self-propelled, in the sense that
in the absence of an external force the filament will slide at a velocity
$v_s$ given by Eq.~\eqref{vs}. The value of $v_s$ increases with increasing motor
stiffness and of course vanishes for $v_0=0$, corresponding  to the vanishing
of the rate of ATP consumption $\Delta\mu$. The sliding velocity $v_s$ is shown
in Fig.~\ref{fig:sliding-vel} as a function of the parameter $\epsilon$ inversely
proportional to the motor stall force for a few values of the maximum number of motors that can bind to the filament. 
A second important effect of motor activity is the discontinuous and hysteretic
response to an external force displayed in Fig.~\ref{hysteresis}. When
$F_\text{ext}=0$ the filament slides at the motor-induced velocity $v_s$. If a small
force $F_\text{ext}>0$ is applied, the filament velocity remains an approximately
linear function of the applied force, but with an effective friction greatly
enhanced by motor binding/unbinding. This enhancement of friction is also termed in the literature as protein friction~\cite{Tawada1991}. At high velocity, only a few motors are attached to the filament and the filament velocity approaches the value it would have in the absence of motors as the applied force is increased beyond a characteristic value. When the external force is ramped
down the filament velocity jumps to the lower branch corresponding to a lower value of the
force, resulting in hysteresis.

\subsection{Fluctuation Dynamics}

We now  examine the dynamics of noise-induced fluctuations about the steady
state by letting  $\delta\dot{x}=\dot{x}-v$,
where $v$ is the steady state velocity, given by the solution of Eq.~\eqref{steady-state}
discussed in the previous section.
  The  dynamics of the
fluctuation $\delta\dot{x}$  is then described by the equation
\begin{eqnarray} 
\label{deltax} 
\zeta\delta\dot{x}= -kN\Delta(v)\delta p_b - kN
p_{bs} \delta\Delta +\eta(t)\;,
\end{eqnarray}
where both $\delta\Delta=[\partial_v\Delta(v)]\delta\dot{x}$ and $\delta p_b(t)$ depend on noise only implicitly through the velocity $\dot{x}$, with
\begin{equation} 
\partial_t\delta
p_b=-\left[\frac{1}{\tau(v)}+\omega_b\right]\delta p_b-p_{bs}(v)\frac{\partial}{\partial v}\left[\frac{1}{\tau(v)}\right] \delta\dot{x}
\label{deltaN}
\end{equation}
The random force  $\eta(t)$ in Eq.~\eqref{deltax}  describes noise  on
the filament, with $\langle\eta(t)\rangle=0$ and
$\langle\eta(t)\eta(t')\rangle=2 B\delta(t-t')$. 
Noise can arise in the system from a variety of sources, including the fluid through which the filament moves and the motor on/off dynamics. For simplicity we assume the spectrum is white, albeit with a non-thermal strength $B$.
By solving Eq.~\eqref{deltaN} with initial condition $\delta p_b(t=0)=0$ and substituting in Eq.~\eqref{deltax},
we obtain a single equation for $\delta\dot{x}$,
\begin{equation}
\left[\zeta+\zeta_a(v)\right]\delta\dot{x}(t)+\omega_0\zeta'_a(v)\int_0^t dt'\
e^{-\Omega(t-t')}\delta\dot{x}(t')= \eta(t) \label{effx}
\end{equation}
where we have introduced  an effective frequency $\Omega(v)=\tau^{-1}(v)+\omega_b$ and active
frictions 
\begin{eqnarray}
&&\zeta_a(v)=kNp_{bs}(v)\partial_v\Delta(v)\\
&&\zeta'_a(v)= kNp_{bs}(v)\Delta(v)\frac{\partial}{\partial v}\left(\frac{1}{\tilde\tau}\right)\;.
\end{eqnarray}
In all the parameters defined above $v$ has to be replaced by the steady state solution obtained in the previous section.
The time scale $\Omega^{-1}$ represent the duration of the cycle of a loaded motor.
Note that $\zeta_a(v=0)=\zeta_{a}$, with $\zeta_a$ given by Eq.~\eqref{zeta-a}. It is  evident from Eq.~\eqref{effx} that motor
dynamics yields a non-Markovian contribution to the friction. 

If we neglect the load dependence of the unbinding rate by letting $\nu=0$, hence $\tau^{-1}=\omega_0$,
then $\zeta_a(v)=\zeta_{a0}=Nrk\ell/v_0$ and $\zeta'_a(v)=0$. In this limit $\langle[\delta x(t)-\delta x(0)]^2\rangle=2 D_{a0} t$ 
and is diffusive 
\emph{at all times}, with an effective diffusion constant 
$D_{a0}=\frac{B}{(\zeta+\zeta_{a0})^2}$. 

 When $\nu$ is finite we obtain
\begin{widetext}
\begin{equation} 
\label{MSD} 
 \langle[\delta x(t)-\delta x(0)]^2 \rangle = 
2D_at +
4D_a\left[\frac{\zeta'_a(v)\omega_0}{[\zeta+\zeta_a(v)]\Omega_a}\right]^2\left(t-\frac{1-e^{-\Omega_a
t}}{\Omega_a}\right)\;, 
\end{equation}
\end{widetext}
where $D_a=B/[\zeta+\zeta_a(v)]^2$ and  $\Omega_a(v)=\Omega(v)+\omega_0\zeta'_a(v)/[\zeta+\zeta_a(v)]$. The characteristic time scale  $\Omega_a^{-1}$ controls the crossover from ballistic behavior for $t \ll \Omega_a^{-1}$ to diffusive behavior for $t \gg \Omega_a^{-1}$. It is determined by the smaller of two time scales: $\Omega^{-1}$, defined after Eq.~\eqref{effx}, that represents the duration of the cycle of a loaded motor, and the active time $(\omega_0\zeta'_a/[\zeta+\zeta_a])^{-1}$ that represents the correlation time for the effect of motor on/off dynamics on the filament. At long times the mean-square displacement  is always diffusive, with an effective diffusion constant
\begin{equation}
D_\text{eff}=D_a\left[1+\left(\frac{\zeta'_a\omega_0}{[\zeta+\zeta_a(v)]\Omega_a}\right)^2\right]
\end{equation}
This result only describes the behavior of the system in the stable region, where the effective friction remains positive. At the onset of negative friction instability  $\zeta+\zeta_a(v)\rightarrow 0$ and the effective diffusivity diverges. In other words the instability is also associated with large fluctuations in he rod's displacements due to the cooperative motor dynamics.

To leading order in $\nu$ the frequency  $\Omega_a$ that controls the crossover to diffusive behavior is simply
$\Omega\simeq\omega_0+\omega_b+{\mathcal O}(\nu^2)$. For non-processive motors such as myosins
$\omega_0 \gg \omega_b$ and $\Omega\sim\omega_0$. 
The effective diffusion constant is given by
\begin{equation}
D_\text{eff}\simeq D_{a}\left[1+ \frac{2\zeta^2\zeta_{a0}}{(\zeta+\zeta_{a0})^3}\left(\frac{v_0\alpha}{\omega_0(1+\epsilon)}\right)^2+{\cal}\left[(v_0\alpha/\omega)^4\right]\right]\;.
\end{equation}
This expression indicates that the enhancement of the diffusion constant comes from the competition of the ballistic motor-driven motion of the filament at speed  $\sim v_0\zeta_{a0}/(\zeta+\zeta_{a0})$ and the randomization of such motion
by the motor on/off dynamics  on time scales $\sim\omega_0^{-1}$. The result is that the filament dynamics is diffusive at long times, but with an enhanced diffusion constant.

Finally, we stress that the correlation function $\langle[\delta x(t)-\delta x(0)]^2 \rangle$ describes the fluctuations about the steady state value $vt$. if we write $x(t)=vt+\delta x(t)$ the actual mean  square displacement of the center of mass of the rod is given by $\langle (x(t)-x(0))^2\rangle= v^2 t^2 + \langle[\delta x(t)-\delta x(0)]^2 \rangle$ and  is ballistic at long times in one dimension due to the mean motion of the rod. In addition, due to nonlinearity of the Langevin equation 
\eqref{xdot} the mean value $\langle x\rangle$ in the presence of noise will in general differ from the steady state solution $vt$ obtained in the absence of noise due to renormalization by fluctuations $\langle F_a(\dot{x},t)\rangle-F_a(v,t)$. These fluctuations are neglected in mean field theory.

\section{Active Filament Dynamics in Two Dimensions}

In two dimensions the coupled translational and rotational dynamics of of the filament is described by Eqs.~\eqref{r-eq} and \eqref{theta-eq}. 
It is convenient to write the instantaneous velocity of the center of the filament in terms of
components longitudinal and transverse to the long axis of the filament,
$\dot{{\bf r}}=V_\Vert {\bf \hat{u}} + V_\perp \bf{\hat {n}}$. Similarly the stretch is written as $\bm\Delta=\Delta_\Vert{\bf \hat{u}}+\Delta_\perp{\bf \hat{n}}$, where (see
Eq.~\eqref{Delta-app})
\begin{subequations} \begin{gather} \Delta_\Vert={\bf
\hat{u}}\cdot\bm\Delta=(V_\Vert+v_m)\tau\;,\label{Delta-par}\\ \Delta_\perp={\bf
\hat{n}}\cdot\bm\Delta=(V_\perp+s_0\dot\theta)\tau\;.\label{Delta-perp}
\end{gather} \end{subequations}
It is then clear that $\Delta_\Vert$ has the same form as in one dimension
\begin{equation} \label{Delta-par2}
\Delta_\Vert=\frac{(V_\Vert-v_0)/\omega_0}{\tilde\tau^{-1}+\epsilon}\;,  \end{equation}
and the mean-field value of the attachment time $\tau$ is
given by
\begin{equation}
\tilde\tau^{-1}(V_\Vert,V_\perp,\dot{\theta})=1+\frac{(V_\Vert-v_0)^2\alpha^2}{(\tilde\tau^{-1}+\epsilon)^2\omega_0^2}+\frac{V_\perp^2\tilde\tau^2\alpha^2}{\omega_0^2}+\frac{L^2\dot\theta^2\tilde\tau^2\alpha^2}{12
\omega_{0}^2}\;, \end{equation}
where we have carried out the average over $s_0$.
Inserting these expressions in Eqs.~\eqref{forceMF} and \eqref{torqueMF}, the
mean field active force and torque exerted by  bound motors on the filament can then be written  as
\begin{subequations}
\begin{gather}
{\bf F}_a=-k N p_b(t) \left[\frac{(V_\Vert-v_0)/\omega_0}{\tilde\tau^{-1}+ \epsilon} \hat{{\bf u}} + V_{\perp}\tau \hat{{\bf n}} \right]\;,\\
T_a=- k N p_b(t) \tau\left[ \frac{L^2\dot{\theta}}{12} + V_{\perp} v_m\tau\right]\;.
\end{gather}
\end{subequations}

\subsection{Steady State and its stability}
The steady state of the motor-driven filament in two dimensions in the absence of noise is characterized by the center of mass velocity ${\bf v}=v_\Vert {\bf \hat{u}} + v_\perp {\bf \hat{n}}$ and angular velocity $\dot{\vartheta}$. In the absence of any external force or torque, $\dot{\vartheta}$ and $v_\perp$ are identically zero, whereas the longitudinal dynamics described by $v_\Vert$ is identical to that obtained in one-dimension: the filament will slide along its long axis at a steady longitudinal velocity 
$v_\Vert=F_p/(\zeta+\zeta_a)$, with $F_p$ and $\zeta_a$ given by Eqs.~\eqref{Fst} and \eqref{zeta-a}, respectively.   

To gain some insight into the stability of the system under application of external forces or torques, we expand ${\bf F}_\text{a}$ and $T_\text{a}$ to linear order in velocities ${\bf v}$ and $\dot{\vartheta}$ as, ${\bf F}_\text{a}({\bf v},\dot{\vartheta})\simeq  {\bf F}_{p}+\left(\frac{\partial {\bf F}_a}{\partial v_\Vert}\right)_{0}v_\Vert + \left(\frac{\partial {\bf F}_a}{\partial v_\perp}\right)_{0}v_\perp+ \left(\frac{\partial {\bf F}_a}{\partial \dot{\vartheta}}\right)_{0}\dot{\vartheta} $, and
$T_a({\bf v},\dot{\vartheta}) \simeq  \left(\frac{\partial  T_a}{\partial v_\Vert}\right)_{0}v_\Vert + \left(\frac{\partial T_a}{\partial v_\perp}\right)_{0}v_\perp + \left(\frac{\partial T_a}{\partial \dot{\vartheta}}\right)_{0}\dot{\vartheta}  $, where ${\bf F}_p={\bf F}_{\text{a},0}=F_p\hat{{\bf u}}$, is the tangential propulsion force due to the motors. The subscript `0' indicates that the expressions are evaluated at ${\bf v}=0$ and $\dot{\vartheta}=0$. This leads to steady state force/velocity and torque/velocity relations of the form
\begin{subequations}
\begin{gather}
	\left(\underline{\underline{{\bm \zeta}}} + \underline{\underline{{\bm \zeta}}}_a\right)\cdot{\bf v}={\bf F}_\text{ext}+ F_p \hat{{\bf u}}\;,\\  
\left(\zeta_\theta+\zeta_{\theta a}\right)\dot{\vartheta}=T_\text{ext}-g_a v_\perp\;,
\end{gather}
\end{subequations}
where we have introduced an active ``momentum" $g_a$ given by $g_a=-\left(\frac{\partial T_a}{\partial v_\perp}\right)_{0}$.
The active contributions to the longitudinal, transverse and rotational friction coefficients are  defined as $\zeta_{\Vert a}=-\hat{{\bf u}}\cdot\left(\frac{\partial {\bf F}_a}{\partial v_\Vert}\right)_{0}$, $\zeta_{\perp a}=-\hat{{\bf n}}\cdot\left(\frac{\partial {\bf F}_a}{\partial v_\perp}\right)_{0}$, and $\zeta_{\theta a}=-\left(\frac{\partial T_a}{\partial \dot{\vartheta}}\right)_{0}$. 
The longitudinal friction coefficient $\zeta_{\Vert a}$ is identical to the active friction $\zeta_a$ given in Eq.~\eqref{zeta-a} for a rod in one dimension, with $\Delta\to\Delta_\Vert$. 
The transverse and rotational friction coefficients are enhanced by motor activity. Their active components are given by
\begin{subequations}
\begin{gather}
\zeta_{\perp a}=\frac{kNr\tau_0}{r+(1-r)\tilde{\tau}_0^{-1}}\\ 
\zeta_{\theta a}=\frac{kNr\tau_0 L^2/12}{r+(1-r)\tilde{\tau}_0^{-1}}\;.
\end{gather}
\end{subequations}
Finally we have, $g_a=\frac{kNr\tau_0v_0\left(\tau_0+\epsilon|\Delta_\Vert^0|\right)}{r+(1-r)\tilde{\tau}_0^{-1}}$.
When the load dependence of the unbinding rate is neglected ($\nu=0$), all friction coefficients are  enhanced by motor activity. When the force/velocity and torque/angular velocity curves are calculated to nonlinear order, we find that the only instability is the negative longitudinal friction instability obtained in one dimension. No instabilities are obtained in the angular dynamics. We expect this will change if we include the semiflexibility of the filament~\cite{Kikuchi2009, Brangwynne2008}.

\subsection{Fluctuations around the steady state} 

We now  examine the dynamics of noise-induced fluctuations about the steady
state by letting  $\delta\dot{{\bf r}}=\dot{{\bf r}}-{\bf v}$ and $\delta\dot{\theta}=\dot{\theta}-\dot{\vartheta}$ where ${\bf v}$ and $\dot{\vartheta}$ are the steady state velocity and angular frequency in the absence of external force and torque.  As noted in the previous section when ${\bf F}_\text{ext}=0$ and $T_\text{ext}=0$, $v_\Vert=v\neq0$, with $v$ given by the solution of Eq.~\eqref{steady-state},  and $v_\perp=\dot{\vartheta}=0$. Projecting velocity fluctuations longitudinal and transverse to the filament, $\delta\dot{{\bf r}}={\bf \hat{u}}\delta V_\Vert+{\bf \hat{n}}\delta V_\perp$, the dynamics of fluctuations is described by the coupled equations,
\begin{subequations}
\begin{gather}
 \label{Vpar}
\left[\zeta_\Vert+\zeta_{\Vert a}(v)\right]\delta V_\Vert=-kN\Delta_\Vert(v)\delta p_b(t) +\eta_\Vert \;,\\
 \label{Vperp} 
\left[\zeta_\perp + \zeta_{\perp a}(v)\right]\delta V_\perp=\eta_\perp  \;,\\
\label{theta} 
\left[\zeta_\theta + \zeta_{\theta a}(v)\right]\delta \dot{\theta}=-kNp_{bs}(v)\tau(v) v_m(v) \delta V_\perp +  \eta_\theta \;, 
\end{gather}
\end{subequations}
with
\begin{equation}
\label{pb} 
\left[\zeta_\theta + \zeta_{\theta a}(v)\right]\delta \dot{p}_b=-\Omega(v)\delta p_b - p_{bs}(v)\frac{\partial}{\partial v}\left[\frac{1}{\tau(v)}\right] \delta V_\Vert \;,
\end{equation}
where the effective frequency $\Omega(v)=\tau^{-1}(v)+\omega_b$ and the longitudinal active friction $\zeta_{\Vert a}(v)$ are as in one dimension,  $\zeta_{\perp a}(v)=kNp_{bs}(v)\tau(v)$ and $\zeta_{\theta a}(v)=kNp_{bs}(v)\tau(v)L^2/12$.  In all the parameters, $v\equiv v_\Vert$ has to be replaced by the steady state solution obtained in one dimension in the absence of external force or torque. 

The time-correlation function of orientational fluctuations, $\Delta\theta(t)=\delta\theta(t)-\delta\theta(0)$, can be calculated from Eqs.~\eqref{Vperp} and \eqref{theta}, with the result
\begin{equation}
\label{ang-cor}
\langle\Delta\theta(t)\Delta\theta(t')\rangle=2 D_{\theta a}\  \text{min}(t,t') \;.
\end{equation}
The effective rotational diffusion constant is  enhanced by the transverse diffusivity and is given by
\begin{equation}
D_{\theta a}(v)=\frac{B_\theta}{\left[\zeta_\theta+\zeta_{\theta a}(v)\right]^2}+\frac{B_\perp/\ell_{p}^2(v)}{\left[\zeta_\perp+\zeta_{\perp a}(v)\right]^2} 
\label{Dtheta-a}
\end{equation}
with $\ell_p(v)= \left[\zeta_\theta + \zeta_{\theta a}(v)\right]/kNp_{bs}(v)\tau(v) v_m(v)$.
Using Eq.~\eqref{ang-cor}, one immediately obtains the angular time-correlation function as~\cite{Han2006},
\begin{equation}
\langle {\bf \hat{u}}(t')\cdot{\bf \hat{u}}(t'') \rangle = e^{-D_{\theta a}
\left| t' -t'' \right|}
\;. \label{ucor}
\end{equation}
The fluctuations in the probability of bound motors are driven by their coupling to the stochastic longitudinal dynamics of the filament. Assuming $\delta p_b(0)=0$, we obtain
\begin{equation}
\label{pbcor}
\langle\delta p_b(t)\delta p_b(t')\rangle=\left(\frac{\zeta_a' \omega_0}{v_p}\right)^2 \frac{B_\Vert}{\Omega_a}
\left[ 
e^{-\Omega_a\left|t-t'\right|} - e^{-\Omega_a\left(t+t'\right)}
\right],
\end{equation}
where $\Omega_a(v)=\Omega(v)+\omega_0\frac{\zeta'_a(v}{\zeta_\Vert+\zeta_{\Vert a}(v)}$, $\zeta'_a(v)= kNp_{bs}(v)\Delta_\Vert(v)\frac{\partial}{\partial v}\left(\frac{1}{\tilde\tau}\right)$, and $v_p(v)=Nk\Delta_\Vert(v)/\left[\zeta_\Vert + \zeta_{\Vert a}(v)\right]$ is a  longitudinal propulsion velocity. Notice that $v_p(v=0)=v_s/p_{bs0}$, with $v_s$ given in Eq.~\eqref{vs}. Finally, we can compute the correlation function of  the fluctuation $\delta\dot{{\bf r}}$ of the filament's position.   In the laboratory frame the dynamics of  $\delta\dot{{\bf r}}$ can be recast in the form of a simple equation, 
\begin{equation}
\delta\dot{{\bf r}}=-v_p \delta p_b(t) \hat{{\bf u}} + \left[\underline{\underline{{\bm \zeta}}} + \underline{\underline{{\bm \zeta}}}^a(v)\right]^{-1}\cdot {\bm \eta}
\end{equation}
Fluctuations in the probability of bound motors do not couple to orientational fluctuations to linear order.  It is then straightforward to calculate  the correlation function of displacement fluctuations, with the result
\begin{widetext}
\begin{eqnarray}
\langle[\delta {\bf r}(t) - \delta {\bf r}(0)]^2\rangle=2D_\text{eff} ~t+ \frac{D_{\Vert a}\zeta_a^{'2} \omega_0^2/\Omega_a^2}{(D_{\theta a}^2-\Omega_a^2)(\zeta_\Vert+\zeta_{\Vert a})^2}\left[-(D_{\theta a}+\Omega_a)\left(1-e^{-2\Omega_a t}\right) 
 +  \frac{4\Omega_a^2}{D_{\theta a}+\Omega_a}\left(1-e^{-(\Omega_a+D_{\theta a}) t}\right)\right] 
 \label{MSD2} 
\end{eqnarray}
\end{widetext}
where effective longitudinal and transverse diffusion constants have been defined as
\begin{subequations}
\begin{gather}
D_{\Vert a}=B_\Vert/[\zeta_\Vert+\zeta_{\Vert a}(v)]^2\;,\\
D_{\perp a}=B_\perp/[\zeta_\perp+\zeta_{\perp a}(v)]^2\;.
\end{gather}
\end{subequations}
Finally, using  ${\bf r}(t)=\delta{\bf r}(t) + \int_0^t \ dt' v \hat{{\bf u}}(t')$, the mean square displacement (MSD) can be written as,  
\begin{equation}
\langle[{\bf r}(t) - {\bf r}(0)]^2\rangle=\langle[\delta {\bf r}(t) - \delta {\bf r}(0)]^2\rangle + \frac{v^2}{D_{\theta a}}\left[t-\frac{1-e^{-D_{\theta a} t}}{D_{\theta a}}\right]\;.
\end{equation}
The MSD is controlled by the interplay of two time scales, 
the rotational diffusion time, $D_{\theta a}^{-1}$, that is decreased by activity as compared to its bare value, $D_\theta^{-1}$, and the time scale $\Omega_a^{-1}$, which is turn controlled by the duration of the motor binding/unbinding cycle. 
If $D_{\theta a}^{-1}\gg \Omega_a^{-1}$, which is indeed the case for actomyosin systems \footnote{A naive estimate for actin-myosin systems (neglecting the load dependence of the unbinding rate) gives $\Omega_{a}^0 \simeq\ 5\ ms^{-1}$ and $D_{\theta a}^0\simeq\ 0.17\ s^{-1}$ for $N=1$. }
then on times $t\gg\Omega_a^{-1}$ the MSD is given by
\begin{equation}
\label{persistent}
\langle[{\bf r}(t) - {\bf r}(0)]^2\rangle=\ 2D_{\text{eff}} t +
 \frac{v^2}{D_{\theta a}}\left[t-\frac{1-e^{-D_{\theta a} t}}{D_{\theta a}}\right]\;,
\end{equation}
with 
\begin{equation}
D_\text{eff}=D_{\Vert a}+D_{\perp a}+\frac{D_{\Vert a}\Omega_a}{D_{\theta a}+\Omega_a}\left(\frac{\zeta'_a\omega_0}{[\zeta_\Vert+\zeta_{\Vert a}(v)]\Omega_a}\right)^2 \;.
\end{equation}
In other words the rod performs a persistent random walk consisting of ballistic segments at speed $v$ randomized by rotational diffusion. The behavior is diffusive both at short and long times, albeit with different diffusion constants, $D_\text{eff}$ and $D_\text{eff}+v^2/(2D_{\theta a})$, respectively. This is indeed the dynamics of a self-propelled rod. If the noise strengths $B_\Vert$, $B_\perp$ and $B_\theta$ are negligible, then Eq. ~\eqref{persistent} reduces to
\begin{equation}
\label{persistent2}
\langle[{\bf r}(t) - {\bf r}(0)]^2\rangle\simeq\ 
 \frac{v^2}{D_{\theta a}}\left[t-\frac{1-e^{-D_{\theta a} t}}{D_{\theta a}}\right]\;.
\end{equation}
and the MSD exhibits a crossover from ballistic behavior for $t\ll D_{\theta a}^{-1}$ to diffusive at long times.

It is worthwhile to note that if one neglects load dependence of unbinding rate by taking $\nu=0$, effective diffusivity at long time is enhanced with, $D_\text{eff}^0=D_{\Vert a}^{0} + D_{\perp a}^{0}  + (v^0)^2/2D_{\theta a}^{0} $, due to the interplay between ballistic motion driven by the tethered motors and rotational diffusion, unlike the situation in one dimension. \\

\section{Summary and Outlook}

We have investigated the dynamics of a single cytoskeletal filament modeled as a rigid rod interacting with tethered motor proteins in  a motility assay in two dimensions.  Motor activity yields both an effective propulsion of the filament along its long axis and a renormalization of all friction coefficients. The longitudinal friction can change sign  leading to an instability in the filament's response to external force, as demonstrated by previous authors~\cite{JulicherProst1995}.  The effective propulsion force and filament velocity in the steady state are calculated in terms of microscopic motor and filament parameters.  

We also considered the fluctuations of the filament displacement about its steady state value and demonstrated that the coupling to the binding/unbinding dynamics of the the motors yields non-Markovian fluctuations and enhanced diffusion.  Future  work will include the stochasticity in the motor displacements and  the semiflexibility of filaments, which is expected to lead to buckling instabilities~\cite{Karpeev2007} and anomalous fluctuations~\cite{TBL2003}.

\acknowledgments

This work was supported at Syracuse by the National Science Foundation under a Materials World Network award 
DMR-0806511 and in Stellenbosch by the National Research Foundation under grant number
UID 67512.
MCM was also partly supported on NSF-DMR-1004789 and NSF-DMR-0705105.   We thank Aparna Baskaran, Lee Boonzaaier and Tannie Liverpool  for illuminating discussions.
Finally, SB and MCM thank the
University of Stellenbosch for hospitality during the completion of
part of this work.

\appendix

\section{Solution of Mean-Field Equation}

Here we discuss the solution of the mean-field equation \eqref{tauMF} for the attachment time $\tau$, For simplicity, we consider the one-dimensional case in detail. The discussion is then easily generalized to two dimensions.
 The mean-field equation for the
residence time $\tau$ is rewritten here for clarity:
\begin{equation}
\tau_{MF}=\omega_u^{-1}(\Delta(\tau_{MF}))\;.
\label{tauMF1d-A} \end{equation}
The solution clearly depends on the form chosen to describe the dependence of the motor unbinding rate on the stretch $\Delta$, in turn given by $\Delta(\tau_{MF})=(\dot{x}-v_0)/(\tau_{MF}^{-1}+\epsilon\omega_0)$. The mean-field equation must be inverted to determine $\tau_{MF}$ as a function of the filament velocity $\dot{x}=v$.
\begin{figure} 
\begin{center} 
\includegraphics[scale=0.7]{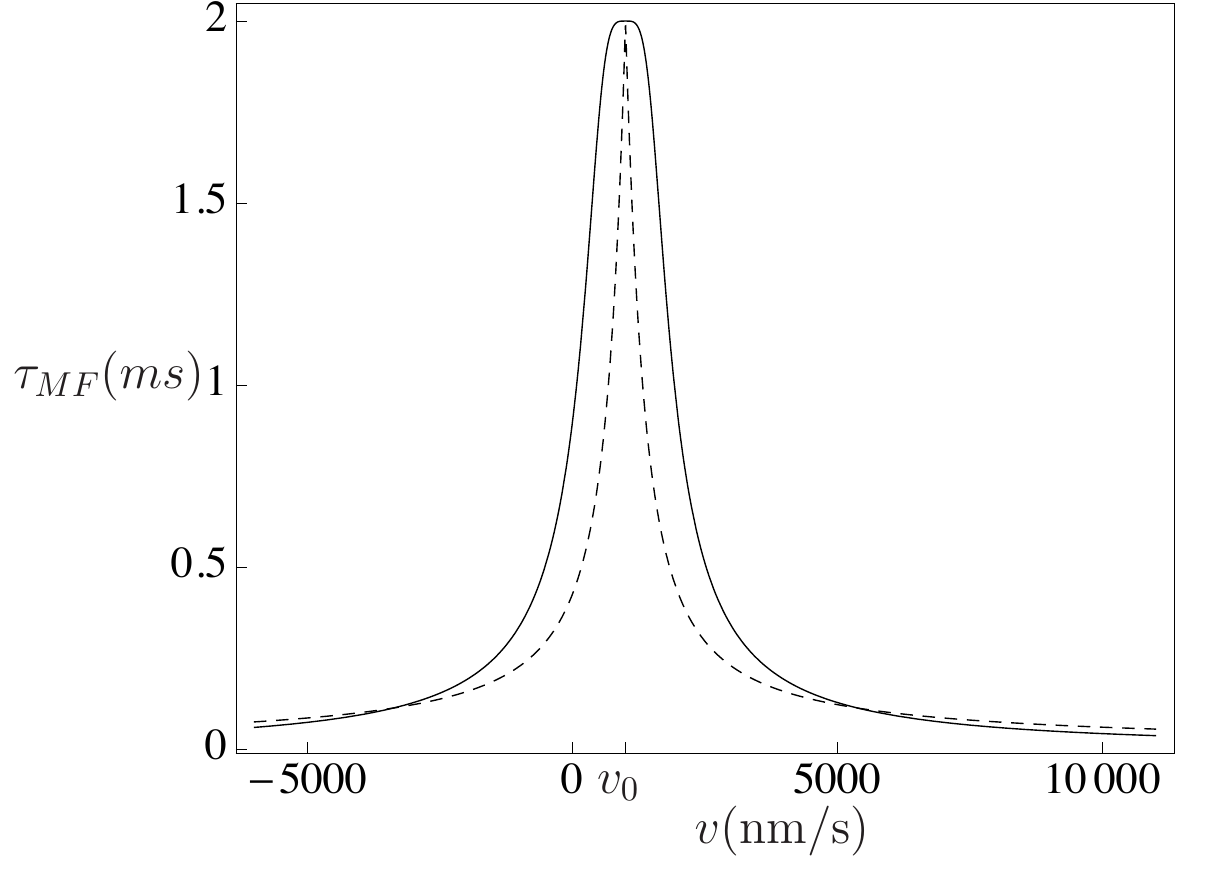}
\end{center} 
\caption{Mean field attachment time $\tau_\text{MF}$ as a function of $v$ for parameter values appropriate for acto-myosin systems:  $v_0=1000\  \text{nm s}^{-1}$, $k=10\ \text{pN
nm}^{-1}$, $f_s=4\ \text{pN}$, $\alpha^{-1}=7.5\ \text{nm}$, $\omega_0=0.5\ \text{(ms)}^{-1}$, $r=0.06$,
corresponding to $\epsilon=5$. The dashed line is the
numerical solution of Eq.~\eqref{tauMF1d-A} obtained using the exponential dependence of the unbinding rate on the stretch. The solid line is obtained using the parabolic ansatz given in Eq.~\eqref{quadratic-omega-A}.}
\label{fig:omegau} 
\end{figure}
For compactness we drop the label `MF'. It is clear that $\tau$ has a maximum at $v=v_0$, where $\tau=\omega_0^{-1}$.
This simply corresponds to the fact that the time a motor protein spends
attached to the actin filament is largest when the motors' tails are
unstretched ($\Delta=0$) and the motors advance at the unloaded
motor velocity, $v_0$.

\begin{figure} 
\begin{center} \includegraphics[scale=0.65]{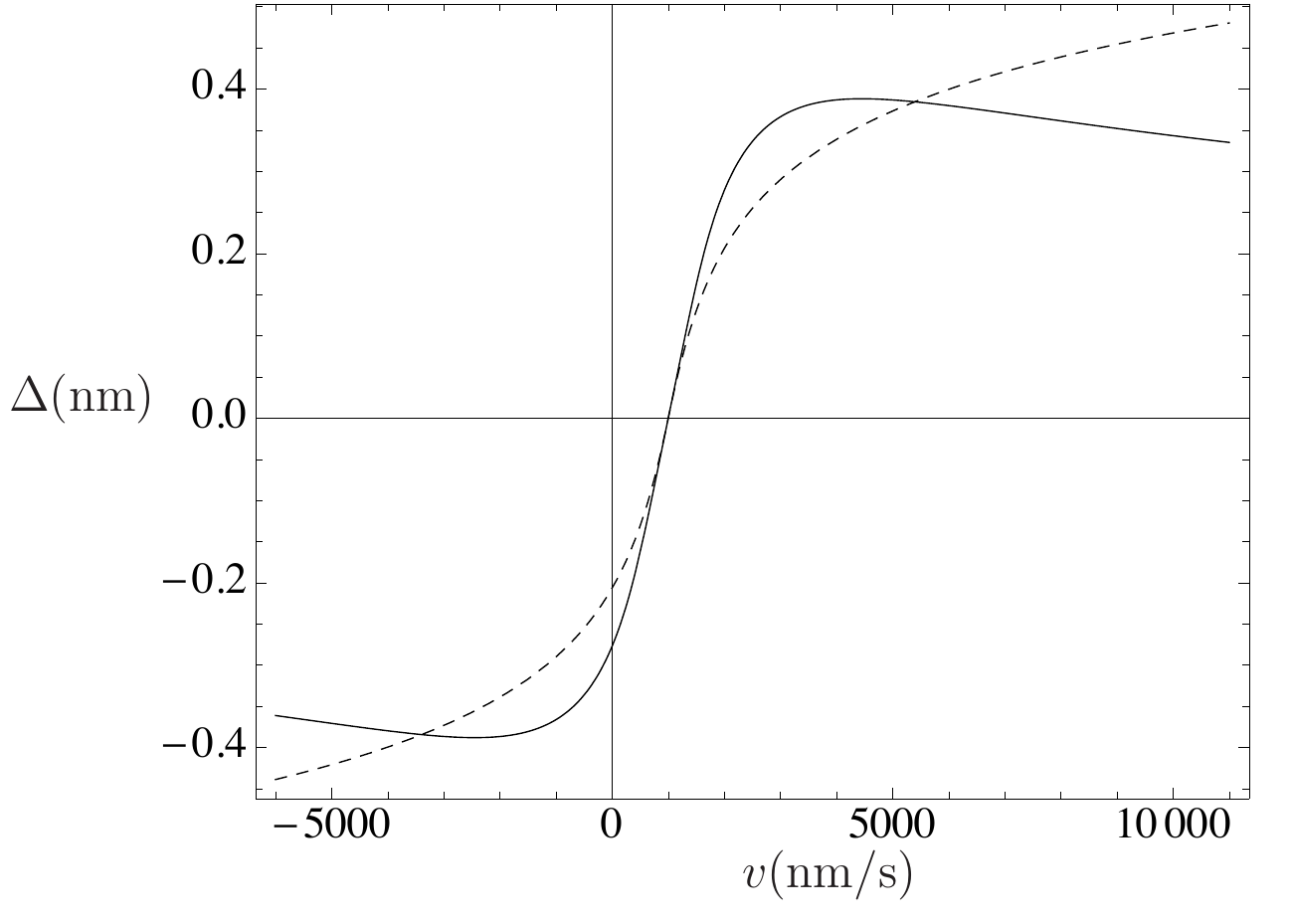}
\end{center} \caption{Stretch $\Delta$ as a function of velocity  $v$ obtained using the mean-field value of the attachment time displayed in Fig.~\ref{fig:omegau}. The parameter values are the same as in Fig.~\ref{fig:omegau}.
The dashed line is  obtained using the exponential dependence of the unbinding rate on the stretch. The solid line is obtained using the parabolic ansatz given in Eq.~\eqref{quadratic-omega-A}.  }
\label{stretch}
\end{figure}

It is convenient to use the dimensionless variable and parameters introduced in the text and write the stretch $\Delta$ as 
\begin{equation}
\Delta=\frac{(u-1)\ell_0}{\tilde\omega_u+1}\;,
\end{equation}
where $u=v/v_0$, $\tilde\omega_u=\omega_u/\omega_0$ and $\ell_0=v_0/\omega_0$.
A form commonly used in the literature is the exponential form $\omega_u(\Delta)=\omega_0e^{\alpha|\Delta|}$, with $\alpha^{-1}$ a characteristic length scale. The dimensionless combination $\alpha\Delta$ can then be written in terms of the parameter $\nu=\alpha\ell=\alpha\ell_0/(1+\epsilon)$ and setting $\nu=0$ corresponds to neglecting the load dependence of the unbinding rate.
The numerical solution of Eq.~\eqref{tauMF1d-A} for the mean  attachment time as a function of $v$ is shown as a dashed line in Fig.~\ref{fig:omegau} for parameter values appropriate for acto-myosin systems. As expected it has a sharp maximum at $v=v_0$. At large $v$ the  attachment time decays logarithmically with velocity. As a result, the stretch is found to saturate at large velocity, as shown by the dashed curve in Fig.~\ref{stretch}. This behavior is unphysical as it does not incorporate the fact that when the stretch exceeds a characteristic value of the order $f_d/k$, the motor head simply detaches, as shown in Fig.~\ref{Fig:vm}. Instead of incorporating this cutoff by hand, we have chosen to use a simple quadratic form for the dependence of the unbinding rate on the stretch, given by
\begin{equation}
\omega_u(\Delta)=\omega_0\left[1+\alpha^2\Delta^2\right]\;.
\label{quadratic-omega-A}
\end{equation}
With this form the mean field equation \eqref{tauMF1d-A} can be solved analytically, although the explicit solution is not terribly informative and will not be given here. 
The resulting  attachment time is shown as a solid line in Fig.~\ref{fig:omegau}.
The quadratic form 
reproduces the sharp maximum of $\tau$ at $v=v_0$ and yields $\tau\sim v^{-3/2}$ at large $v$. The stretch then decays with velocity, as shown in Fig.~\ref{stretch}.

\bibliography{references}

\begin{thebibliography}{33}
\expandafter\ifx\csname natexlab\endcsname\relax\def\natexlab#1{#1}\fi
\expandafter\ifx\csname bibnamefont\endcsname\relax
  \def\bibnamefont#1{#1}\fi
\expandafter\ifx\csname bibfnamefont\endcsname\relax
  \def\bibfnamefont#1{#1}\fi
\expandafter\ifx\csname citenamefont\endcsname\relax
  \def\citenamefont#1{#1}\fi
\expandafter\ifx\csname url\endcsname\relax
  \def\url#1{\texttt{#1}}\fi
\expandafter\ifx\csname urlprefix\endcsname\relax\def\urlprefix{URL }\fi
\providecommand{\bibinfo}[2]{#2}
\providecommand{\eprint}[2][]{\url{#2}}

\bibitem[{\citenamefont{Schaller et~al.}(2010)\citenamefont{Schaller, Weber,
  Semmrich, Frey, and Bausch}}]{Schaller2010}
\bibinfo{author}{\bibfnamefont{V.}~\bibnamefont{Schaller}},
  \bibinfo{author}{\bibfnamefont{C.}~\bibnamefont{Weber}},
  \bibinfo{author}{\bibfnamefont{C.}~\bibnamefont{Semmrich}},
  \bibinfo{author}{\bibfnamefont{E.}~\bibnamefont{Frey}}, \bibnamefont{and}
  \bibinfo{author}{\bibfnamefont{A.~R.} \bibnamefont{Bausch}},
  \bibinfo{journal}{Nature} \textbf{\bibinfo{volume}{467}}, \bibinfo{pages}{73}
  (\bibinfo{year}{2010}).

\bibitem[{\citenamefont{Butt et~al.}(2010)\citenamefont{Butt, Mufti, Humayun,
  Rosenthal, Khan, Khan, and Molloy}}]{Butt2010}
\bibinfo{author}{\bibfnamefont{T.}~\bibnamefont{Butt}},
  \bibinfo{author}{\bibfnamefont{T.}~\bibnamefont{Mufti}},
  \bibinfo{author}{\bibfnamefont{A.}~\bibnamefont{Humayun}},
  \bibinfo{author}{\bibfnamefont{P.~B.} \bibnamefont{Rosenthal}},
  \bibinfo{author}{\bibfnamefont{S.}~\bibnamefont{Khan}},
  \bibinfo{author}{\bibfnamefont{S.}~\bibnamefont{Khan}}, \bibnamefont{and}
  \bibinfo{author}{\bibfnamefont{J.~E.} \bibnamefont{Molloy}},
  \bibinfo{journal}{J. Biol. Chem.} \textbf{\bibinfo{volume}{285}},
  \bibinfo{pages}{4964} (\bibinfo{year}{2010}).

\bibitem[{\citenamefont{Copeland and Weibel}(2009)}]{Weibel2009}
\bibinfo{author}{\bibfnamefont{M.~F.} \bibnamefont{Copeland}} \bibnamefont{and}
  \bibinfo{author}{\bibfnamefont{D.~B.} \bibnamefont{Weibel}},
  \bibinfo{journal}{Soft Matter} \textbf{\bibinfo{volume}{5}},
  \bibinfo{pages}{1174} (\bibinfo{year}{2009}).

\bibitem[{\citenamefont{Riveline et~al.}(1998)\citenamefont{Riveline, Ott,
  J\"ulicher, Winkelmann, Cardoso, Lacap\`ere, Magn\'usd\'ottir, Viovy,
  Gorre-Talini, and Prost}}]{Riveline1998}
\bibinfo{author}{\bibfnamefont{D.}~\bibnamefont{Riveline}},
  \bibinfo{author}{\bibfnamefont{A.}~\bibnamefont{Ott}},
  \bibinfo{author}{\bibfnamefont{F.}~\bibnamefont{J\"ulicher}},
  \bibinfo{author}{\bibfnamefont{D.~A.} \bibnamefont{Winkelmann}},
  \bibinfo{author}{\bibfnamefont{O.}~\bibnamefont{Cardoso}},
  \bibinfo{author}{\bibfnamefont{J.-J.} \bibnamefont{Lacap\`ere}},
  \bibinfo{author}{\bibfnamefont{S.}~\bibnamefont{Magn\'usd\'ottir}},
  \bibinfo{author}{\bibfnamefont{J.~L.} \bibnamefont{Viovy}},
  \bibinfo{author}{\bibfnamefont{L.}~\bibnamefont{Gorre-Talini}},
  \bibnamefont{and} \bibinfo{author}{\bibfnamefont{J.}~\bibnamefont{Prost}},
  \bibinfo{journal}{Eur. Biophys. J.} \textbf{\bibinfo{volume}{27}},
  \bibinfo{pages}{403} (\bibinfo{year}{1998}).

\bibitem[{\citenamefont{Gu{\'e}rin
  et~al.}(2010{\natexlab{a}})\citenamefont{Gu{\'e}rin, Prost, Martin, and
  Joanny}}]{GuerinReview2010}
\bibinfo{author}{\bibfnamefont{T.}~\bibnamefont{Gu{\'e}rin}},
  \bibinfo{author}{\bibfnamefont{J.}~\bibnamefont{Prost}},
  \bibinfo{author}{\bibfnamefont{P.}~\bibnamefont{Martin}}, \bibnamefont{and}
  \bibinfo{author}{\bibfnamefont{J.-F.} \bibnamefont{Joanny}},
  \bibinfo{journal}{Curr. Op. Cell Biol.} \textbf{\bibinfo{volume}{22}},
  \bibinfo{pages}{14} (\bibinfo{year}{2010}{\natexlab{a}}).

\bibitem[{\citenamefont{J{\"u}licher and Prost}(1997)}]{JulicherProst1997}
\bibinfo{author}{\bibfnamefont{F.}~\bibnamefont{J{\"u}licher}}
  \bibnamefont{and} \bibinfo{author}{\bibfnamefont{J.}~\bibnamefont{Prost}},
  \bibinfo{journal}{Phys. Rev. Lett.} \textbf{\bibinfo{volume}{78}},
  \bibinfo{pages}{4510} (\bibinfo{year}{1997}),
  \urlprefix\url{http://link.aps.org/doi/10.1103/PhysRevLett.78.4510}.

\bibitem[{\citenamefont{Grill et~al.}(2005)\citenamefont{Grill, Kruse, and
  J{\"u}licher}}]{Grill2005}
\bibinfo{author}{\bibfnamefont{S.~W.} \bibnamefont{Grill}},
  \bibinfo{author}{\bibfnamefont{K.}~\bibnamefont{Kruse}}, \bibnamefont{and}
  \bibinfo{author}{\bibfnamefont{F.}~\bibnamefont{J{\"u}licher}},
  \bibinfo{journal}{Phys. Rev. Lett.} \textbf{\bibinfo{volume}{94}},
  \bibinfo{pages}{108104} (\bibinfo{year}{2005}),
  \urlprefix\url{http://dx.doi.org/10.1103/PhysRevLett.94.108104}.

\bibitem[{\citenamefont{G{\"u}nther and Kruse}(2007)}]{GuntherKruse2007}
\bibinfo{author}{\bibfnamefont{S.}~\bibnamefont{G{\"u}nther}} \bibnamefont{and}
  \bibinfo{author}{\bibfnamefont{K.}~\bibnamefont{Kruse}},
  \bibinfo{journal}{New J. Phys.} \textbf{\bibinfo{volume}{9}},
  \bibinfo{pages}{417} (\bibinfo{year}{2007}),
  \urlprefix\url{http://iopscience.iop.org/1367-2630/9/11/417/}.

\bibitem[{\citenamefont{Vilfan and Frey}(2005)}]{VilfanFrey2005}
\bibinfo{author}{\bibfnamefont{A.}~\bibnamefont{Vilfan}} \bibnamefont{and}
  \bibinfo{author}{\bibfnamefont{E.}~\bibnamefont{Frey}},
  \bibinfo{journal}{Journal of Physics: Condensed Matter}
  \textbf{\bibinfo{volume}{17}}, \bibinfo{pages}{S3901} (\bibinfo{year}{2005}),
  \urlprefix\url{http://stacks.iop.org/0953-8984/17/i=47/a=018}.

\bibitem[{\citenamefont{Camalet and J{\"u}licher}(2000)}]{CamaletJulicher2000}
\bibinfo{author}{\bibfnamefont{S.}~\bibnamefont{Camalet}} \bibnamefont{and}
  \bibinfo{author}{\bibfnamefont{F.}~\bibnamefont{J{\"u}licher}},
  \bibinfo{journal}{New Journal of Physics} \textbf{\bibinfo{volume}{2}},
  \bibinfo{pages}{24} (\bibinfo{year}{2000}),
  \urlprefix\url{http://stacks.iop.org/1367-2630/2/i=1/a=324}.

\bibitem[{\citenamefont{J{\"u}licher and Prost}(1995)}]{JulicherProst1995}
\bibinfo{author}{\bibfnamefont{F.}~\bibnamefont{J{\"u}licher}}
  \bibnamefont{and} \bibinfo{author}{\bibfnamefont{J.}~\bibnamefont{Prost}},
  \bibinfo{journal}{Phys. Rev. Lett.} \textbf{\bibinfo{volume}{75}},
  \bibinfo{pages}{2618} (\bibinfo{year}{1995}),
  \urlprefix\url{http://link.aps.org/doi/10.1103/PhysRevLett.75.2618}.

\bibitem[{\citenamefont{Badoual et~al.}(2002)\citenamefont{Badoual,
  J{\"u}licher, and Prost}}]{Badoual2002}
\bibinfo{author}{\bibfnamefont{M.}~\bibnamefont{Badoual}},
  \bibinfo{author}{\bibfnamefont{F.}~\bibnamefont{J{\"u}licher}},
  \bibnamefont{and} \bibinfo{author}{\bibfnamefont{J.}~\bibnamefont{Prost}},
  \bibinfo{journal}{Proc. Natl. Acad. Sci. USA} \textbf{\bibinfo{volume}{99}},
  \bibinfo{pages}{6696} (\bibinfo{year}{2002}).

\bibitem[{\citenamefont{Pla{\c c}ais et~al.}(2009)\citenamefont{Pla{\c c}ais,
  Balland, Gu{\'e}rin, Joanny, and Martin}}]{Placais2009}
\bibinfo{author}{\bibfnamefont{P.~Y.} \bibnamefont{Pla{\c c}ais}},
  \bibinfo{author}{\bibfnamefont{M.}~\bibnamefont{Balland}},
  \bibinfo{author}{\bibfnamefont{T.}~\bibnamefont{Gu{\'e}rin}},
  \bibinfo{author}{\bibfnamefont{J.-F.} \bibnamefont{Joanny}},
  \bibnamefont{and} \bibinfo{author}{\bibfnamefont{P.}~\bibnamefont{Martin}},
  \bibinfo{journal}{Phys. Rev. Lett.} \textbf{\bibinfo{volume}{103}},
  \bibinfo{pages}{158102} (\bibinfo{year}{2009}),
  \urlprefix\url{http://link.aps.org/doi/10.1103/PhysRevLett.103.158102}.

\bibitem[{\citenamefont{Gibbons et~al.}(2001)\citenamefont{Gibbons, Chauwin,
  Desp\'osito, and Jos{\'e}}}]{Gibbons2001}
\bibinfo{author}{\bibfnamefont{F.}~\bibnamefont{Gibbons}},
  \bibinfo{author}{\bibfnamefont{J.~F.} \bibnamefont{Chauwin}},
  \bibinfo{author}{\bibfnamefont{M.}~\bibnamefont{Desp\'osito}},
  \bibnamefont{and} \bibinfo{author}{\bibfnamefont{J.~V.}
  \bibnamefont{Jos{\'e}}}, \bibinfo{journal}{Biophys. J.}
  \textbf{\bibinfo{volume}{80}}, \bibinfo{pages}{2515} (\bibinfo{year}{2001}).

\bibitem[{\citenamefont{Kraikivski et~al.}(2006)\citenamefont{Kraikivski,
  Lipowsky, and Kierfeld}}]{Kraikivski2006}
\bibinfo{author}{\bibfnamefont{P.}~\bibnamefont{Kraikivski}},
  \bibinfo{author}{\bibfnamefont{R.}~\bibnamefont{Lipowsky}}, \bibnamefont{and}
  \bibinfo{author}{\bibfnamefont{J.}~\bibnamefont{Kierfeld}},
  \bibinfo{journal}{Phys. Rev. Lett.} \textbf{\bibinfo{volume}{96}},
  \bibinfo{pages}{258103} (\bibinfo{year}{2006}),
  \urlprefix\url{http://link.aps.org/doi/10.1103/PhysRevLett.96.258103}.

\bibitem[{\citenamefont{Brokaw}(1975)}]{Brokaw1975}
\bibinfo{author}{\bibfnamefont{C.~J.} \bibnamefont{Brokaw}},
  \bibinfo{journal}{Proc. Natl. Acad. Sci. USA} \textbf{\bibinfo{volume}{72}},
  \bibinfo{pages}{3102} (\bibinfo{year}{1975}).

\bibitem[{\citenamefont{Vilfan et~al.}(1999)\citenamefont{Vilfan, Frey, and
  Schwabl}}]{Vilfan1999}
\bibinfo{author}{\bibfnamefont{A.}~\bibnamefont{Vilfan}},
  \bibinfo{author}{\bibfnamefont{E.}~\bibnamefont{Frey}}, \bibnamefont{and}
  \bibinfo{author}{\bibfnamefont{F.}~\bibnamefont{Schwabl}},
  \bibinfo{journal}{Europhys. Lett.} \textbf{\bibinfo{volume}{283}},
  \bibinfo{pages}{45} (\bibinfo{year}{1999}).

\bibitem[{\citenamefont{Hexner and Kafri}(2009)}]{Kafri2009}
\bibinfo{author}{\bibfnamefont{D.}~\bibnamefont{Hexner}} \bibnamefont{and}
  \bibinfo{author}{\bibfnamefont{Y.}~\bibnamefont{Kafri}},
  \bibinfo{journal}{Phys Biol} \textbf{\bibinfo{volume}{6}},
  \bibinfo{pages}{036016} (\bibinfo{year}{2009}),
  \urlprefix\url{http://iopscience.iop.org/1478-3975/6/3/036016}.

\bibitem[{\citenamefont{Gu{\'e}rin
  et~al.}(2010{\natexlab{b}})\citenamefont{Gu{\'e}rin, Prost, and
  Joanny}}]{Guerin2010}
\bibinfo{author}{\bibfnamefont{T.}~\bibnamefont{Gu{\'e}rin}},
  \bibinfo{author}{\bibfnamefont{J.}~\bibnamefont{Prost}}, \bibnamefont{and}
  \bibinfo{author}{\bibfnamefont{J.-F.} \bibnamefont{Joanny}},
  \bibinfo{journal}{Phys. Rev. Lett.} \textbf{\bibinfo{volume}{104}},
  \bibinfo{pages}{248102} (\bibinfo{year}{2010}{\natexlab{b}}),
  \urlprefix\url{http://prl.aps.org/abstract/PRL/v104/i24/e248102}.

\bibitem[{\citenamefont{Huxley}(1957)}]{Huxley1957}
\bibinfo{author}{\bibfnamefont{A.~F.} \bibnamefont{Huxley}},
  \bibinfo{journal}{Prog. Biophys. Chem.} \textbf{\bibinfo{volume}{7}},
  \bibinfo{pages}{255} (\bibinfo{year}{1957}).

\bibitem[{\citenamefont{Vilfan}(2009)}]{Vilfan2009}
\bibinfo{author}{\bibfnamefont{A.}~\bibnamefont{Vilfan}},
  \bibinfo{journal}{Biophys. J.} \textbf{\bibinfo{volume}{1130Ð1137}},
  \bibinfo{pages}{2515} (\bibinfo{year}{2009}).

\bibitem[{\citenamefont{van Teeffelen and L{\"o}wen}(2008)}]{Teeffelen2008}
\bibinfo{author}{\bibfnamefont{S.}~\bibnamefont{van Teeffelen}}
  \bibnamefont{and}
  \bibinfo{author}{\bibfnamefont{H.}~\bibnamefont{L{\"o}wen}},
  \bibinfo{journal}{Phys. Rev. E} \textbf{\bibinfo{volume}{78}},
  \bibinfo{pages}{020101} (\bibinfo{year}{2008}),
  \urlprefix\url{http://pre.aps.org/abstract/PRE/v78/i2/e020101}.

\bibitem[{\citenamefont{Baskaran and Marchetti}(2008)}]{BaskaranMarchetti2008}
\bibinfo{author}{\bibfnamefont{A.}~\bibnamefont{Baskaran}} \bibnamefont{and}
  \bibinfo{author}{\bibfnamefont{M.~C.} \bibnamefont{Marchetti}},
  \bibinfo{journal}{Phys. Rev. Lett.} \textbf{\bibinfo{volume}{101}},
  \bibinfo{pages}{268101} (\bibinfo{year}{2008}).

\bibitem[{\citenamefont{Svoboda and Block}(1994)}]{Svoboda1994}
\bibinfo{author}{\bibfnamefont{K.}~\bibnamefont{Svoboda}} \bibnamefont{and}
  \bibinfo{author}{\bibfnamefont{S.~M.} \bibnamefont{Block}},
  \bibinfo{journal}{Cell} \textbf{\bibinfo{volume}{77}}, \bibinfo{pages}{773}
  (\bibinfo{year}{1994}).

\bibitem[{\citenamefont{Parmeggiani et~al.}(2001)\citenamefont{Parmeggiani,
  J{\"u}licher, Peliti, and Prost}}]{Parmeggiani2001}
\bibinfo{author}{\bibfnamefont{A.}~\bibnamefont{Parmeggiani}},
  \bibinfo{author}{\bibfnamefont{F.}~\bibnamefont{J{\"u}licher}},
  \bibinfo{author}{\bibfnamefont{L.}~\bibnamefont{Peliti}}, \bibnamefont{and}
  \bibinfo{author}{\bibfnamefont{J.}~\bibnamefont{Prost}},
  \bibinfo{journal}{Europhys. Lett.} \textbf{\bibinfo{volume}{56}},
  \bibinfo{pages}{603} (\bibinfo{year}{2001}), \eprint{cond-mat/0109187v1},
  \urlprefix\url{http://arxiv.org/abs/cond-mat/0109187v1}.

\bibitem[{\citenamefont{Visscher et~al.}(1999)\citenamefont{Visscher,
  Schnitzer, and Block}}]{Visscher1999}
\bibinfo{author}{\bibfnamefont{K.}~\bibnamefont{Visscher}},
  \bibinfo{author}{\bibfnamefont{M.~J.} \bibnamefont{Schnitzer}},
  \bibnamefont{and} \bibinfo{author}{\bibfnamefont{S.~M.} \bibnamefont{Block}},
  \bibinfo{journal}{Nature} \textbf{\bibinfo{volume}{400}},
  \bibinfo{pages}{184} (\bibinfo{year}{1999}),
  \urlprefix\url{http://www.nature.com/nature/journal/v400/n6740/abs/400184a0.html}.

\bibitem[{\citenamefont{Howard}(2001)}]{Howard2001}
\bibinfo{author}{\bibfnamefont{J.}~\bibnamefont{Howard}},
  \emph{\bibinfo{title}{{Mechanics of Motor Proteins and the Cytoskeleton}}}
  (\bibinfo{publisher}{Sinauer Associates}, \bibinfo{year}{2001}), ISBN
  \bibinfo{isbn}{0878933344},
  \urlprefix\url{http://www.amazon.com/exec/obidos/redirect?tag=citeulike07-20\&path=ASIN/0878933344}.

\bibitem[{\citenamefont{Tawada and Sekimoto}(1991)}]{Tawada1991}
\bibinfo{author}{\bibfnamefont{K.}~\bibnamefont{Tawada}} \bibnamefont{and}
  \bibinfo{author}{\bibfnamefont{K.}~\bibnamefont{Sekimoto}},
  \bibinfo{journal}{Journal of Theoretical Biology}
  \textbf{\bibinfo{volume}{150}}, \bibinfo{pages}{193 } (\bibinfo{year}{1991}),
  ISSN \bibinfo{issn}{0022-5193},
  \urlprefix\url{http://www.sciencedirect.com/science/article/B6WMD-4KDGR4D-5/2/906e9dbabbba82beff6bc6e5f982d9bd}.

\bibitem[{\citenamefont{Kikuchi et~al.}(2009)\citenamefont{Kikuchi, Ehrlicher,
  Koch, K{\"a}s, Ramaswamy, and Rao}}]{Kikuchi2009}
\bibinfo{author}{\bibfnamefont{N.}~\bibnamefont{Kikuchi}},
  \bibinfo{author}{\bibfnamefont{A.}~\bibnamefont{Ehrlicher}},
  \bibinfo{author}{\bibfnamefont{D.}~\bibnamefont{Koch}},
  \bibinfo{author}{\bibfnamefont{J.~A.} \bibnamefont{K{\"a}s}},
  \bibinfo{author}{\bibfnamefont{S.}~\bibnamefont{Ramaswamy}},
  \bibnamefont{and} \bibinfo{author}{\bibfnamefont{M.}~\bibnamefont{Rao}},
  \bibinfo{journal}{Proceedings of the National Academy of Sciences}
  \textbf{\bibinfo{volume}{106}}, \bibinfo{pages}{19776}
  (\bibinfo{year}{2009}).

\bibitem[{\citenamefont{Brangwynne et~al.}(2008)\citenamefont{Brangwynne,
  Koenderink, MacKintosh, and Weitz}}]{Brangwynne2008}
\bibinfo{author}{\bibfnamefont{C.~P.} \bibnamefont{Brangwynne}},
  \bibinfo{author}{\bibfnamefont{G.~H.} \bibnamefont{Koenderink}},
  \bibinfo{author}{\bibfnamefont{F.~C.} \bibnamefont{MacKintosh}},
  \bibnamefont{and} \bibinfo{author}{\bibfnamefont{D.~A.} \bibnamefont{Weitz}},
  \bibinfo{journal}{Phys. Rev. Lett.} \textbf{\bibinfo{volume}{100}},
  \bibinfo{pages}{118104} (\bibinfo{year}{2008}).

\bibitem[{\citenamefont{Han et~al.}(2006)\citenamefont{Han, Alsayed, Nobili,
  Zhang, Lubensky, and Yodh}}]{Han2006}
\bibinfo{author}{\bibfnamefont{Y.}~\bibnamefont{Han}},
  \bibinfo{author}{\bibfnamefont{A.}~\bibnamefont{Alsayed}},
  \bibinfo{author}{\bibfnamefont{M.}~\bibnamefont{Nobili}},
  \bibinfo{author}{\bibfnamefont{J.}~\bibnamefont{Zhang}},
  \bibinfo{author}{\bibfnamefont{T.~C.} \bibnamefont{Lubensky}},
  \bibnamefont{and} \bibinfo{author}{\bibfnamefont{A.~G.} \bibnamefont{Yodh}},
  \bibinfo{journal}{Science} \textbf{\bibinfo{volume}{314}},
  \bibinfo{pages}{626} (\bibinfo{year}{2006}),
  \urlprefix\url{http://www.sciencemag.org/content/314/5799/626.short}.

\bibitem[{\citenamefont{Karpeev et~al.}(2007)\citenamefont{Karpeev, Aranson,
  Tsimring, and Kaper}}]{Karpeev2007}
\bibinfo{author}{\bibfnamefont{D.}~\bibnamefont{Karpeev}},
  \bibinfo{author}{\bibfnamefont{I.~S.} \bibnamefont{Aranson}},
  \bibinfo{author}{\bibfnamefont{L.~S.} \bibnamefont{Tsimring}},
  \bibnamefont{and} \bibinfo{author}{\bibfnamefont{H.~G.} \bibnamefont{Kaper}},
  \bibinfo{journal}{Phys. Rev. E} \textbf{\bibinfo{volume}{76}},
  \bibinfo{pages}{051905} (\bibinfo{year}{2007}).

\bibitem[{\citenamefont{Liverpool}(2003)}]{TBL2003}
\bibinfo{author}{\bibfnamefont{T.~B.} \bibnamefont{Liverpool}},
  \bibinfo{journal}{Phys. Rev. E} \textbf{\bibinfo{volume}{67}},
  \bibinfo{pages}{031909} (\bibinfo{year}{2003}),
  \urlprefix\url{http://pre.aps.org/abstract/PRE/v67/i3/e031909}.

\end{thebibliography}

\end{document}